\documentstyle[12pt,ssi]{article}

\input epsf 

\newcommand{\rs}{\mbox{${\rm \sqrt{s} \;}$}}
\newcommand{\mxtwo}{\mbox{${\rm M^2_X \;}$}} 
\newcommand{\stot}{\mbox{${\rm \sigma_{tot}(s)}$}} 
\newcommand{\mx}{\mbox{${\rm M_X \;}$}} 
\newcommand{\gevtwom}{\mbox{\rm Ge$V^2$}}
\newcommand{\gevmtwom}{\mbox{\rm Ge$V^{-2}$}}
\newcommand{\qtwom}{\mbox{$Q^2 \;$}}

\newcommand{\xbj}{\mbox{$\rm x_{Bj} \;$}}
 \newcommand {\pomer} {\mbox{ ${I\hspace{-0.25em}P}$ }} 
 \newcommand {\pom}  {I\hspace{-0.15em}P} 
\newcommand{\xpom} {\mbox{${\rm x_{_{\pom}}\;}$}}

 \newcommand{\xl} {${\rm x_L \;}$}
\newcommand{\ra} {\rightarrow} 
\newcommand{\ram} {$\rightarrow$}
\newcommand{\gpm} {${\rm \gamma p}$}

\newcommand{\alphapom} {\rm \alpha_{_{\pom}}} 
\newcommand{\alphapombar} {\rm \overline{\alphapom}} 
\newcommand{\ftwodthree} {${\rm F_2^{D(3)}(Q^2,\beta,\xpom)}$}

\newcommand{\qqbar} {${\rm \; q\bar{q}\;}$} 
 
\def\Wgp{W_{\gamma p}}
\def\Wgpm {\mbox {${\rm W_{\gamma p \;}}$}}
\newcommand{\rb} {\rm\boldmath}
\newcommand{\stgp} {\mbox{${\rm \sigma_{tot}^{\gamma p}\;}$}}
\def\deta{$\Delta\eta \;$}

\newcommand{\pf} {\protect\footnotesize}
\newcommand{\bi} {\begin{itemize}} \newcommand{\ei} {\end{itemize}}
\newcommand{\be} {\begin{equation}} \newcommand{\ee} {\end{equation}}
\newcommand{\bc} {\begin{center}} \newcommand{\ec} {\end{center}}
\newcommand{\noi} {\noindent} 
\newcommand{\nl} {\newline} 
\newcommand{\lm} {\vspace{-1.cm}\caption} 
\newcommand{\bfm}{\begin{figure}[htb]\vspace{-0.6cm}}
\newcommand{\pl} {Phys. Lett.} 
\newcommand{\zp} {Z. Phys.} 
\newcommand{\prev} {Phys. Rev.}
\newcommand{\prep} {Phys. Rep.}

\newcommand{\spj} {Sov. Phys. JEPT}
\newcommand{\prl} {Phys. Rev. Lett.}
\newcommand{\np} {Nucl. Phys.}
\newcommand{\zcol}{ZEUS Collaboration, M. Derrick et al.}
\newcommand{\hcol}{H1 Collaboration, F. Abe et al.}
\newcommand{\hcola}{H1 Collaboration, S. Aid et al.}
   
\begin{document}

\title{ Diffraction at HERA}

\author{Nicol\`{o} Cartiglia \\
Columbia University, Nevis Laboratories, \\
136 South Broadway, Irvington N.Y., 10533 USA \\[0.4cm]
Representing the {\large H1} and {\large ZEUS} Collaborations
}
\maketitle
\begin{abstract}
\noi Recent results on diffraction at HERA, as measured by the H1 and ZEUS collaborations, are reviewed. Results on the photon-proton total hadronic cross section, on vector meson production both at small and large photon virtuality and on photon diffraction are presented. The experimental signature of diffraction at HERA, as well as the selection methods used by the two collaborations are explained.
\end{abstract}

\tableofcontents
\newpage
\section{Introduction}
 Photon-proton collisions have been extensively studied in fixed target
experiments up to centre of mass energies, \Wgpm, of about 20 GeV,  using both
real and virtual photons. At the HERA collider at DESY, 820 GeV protons collide
with 27.5 GeV electrons or positrons. The HERA physics program is very rich,
ranging from  non-perturbative to perturbative QCD, heavy-flavour physics and
to the measurement of the quark and
 gluon densities in the proton and in the photon. Two general purpose detectors,
H1\cite{h1} and ZEUS\cite{zeus} , operate at HERA and  are instrumented with high resolution
calorimeters and tracking chambers.

The results  presented here have been obtained using  
data collected during 1994 and 1995, for a total of about ${\rm 9 ~pb^{-1}}$. More detailed presentations on individual subjects  can be found in many proceedings\cite{ref:peppe} .

\section{Diffraction and total cross section}
\label{sec:regge}
Historically, hadronic diffraction processes and total cross sections have been described using the concept of `pomeron exchange'.
The simplest way to introduce the concept of pomeron is within the framework of  Regge theory\cite{ref:col,ref:pearl} . 
Consider the example shown in
Fig.~\ref{fig:traj}: ${\pi^- p \rightarrow \pi^o n}$ where t is the 4-momentum transfer. According to quantum numbers conservation, this reaction might
happen via the exchange of a virtual  ${ \rho^0, \;a_2, \; g }$ hadron. If  the values of the masses and spins of these particles are plotted on the right hand side of the spin-t plane (where t is positive), they lie almost on a straight line determining a 'trajectory' of particles.
The general expression for a straight line  trajectory is:
$${\rm \alpha(t) = \alpha(0)+\alpha^{\prime}\cdot t},$$
\noi where  ${\alpha(0)}$ is the intercept and ${\alpha^{\prime}}$ the slope. 
\noi The most important trajectories are approximately linear with a universal slope
${\rm \alpha^{\prime}~=~0.9}$ \gevtwom; the first
particle on a trajectory gives the name  to the trajectory itself (in the
above example the $\rho$ trajectory is exchanged). 
 Regge theory predicts that the properties  of a t-channel reaction (that happens on the left hand side of the spin-t plane, where t is negative, via the exchange of off mass shell particles), ${\pi^- p \rightarrow \pi^o n}$ for example,  are determined by the parameters of the trajectory formed by the exchanged particles on the right hand side of the spin-t plane (the $\rho$ trajectory in the case above). 

\bfm
\begin{center}
\leavevmode
\hbox{%
\epsfxsize =4.in
 \epsffile{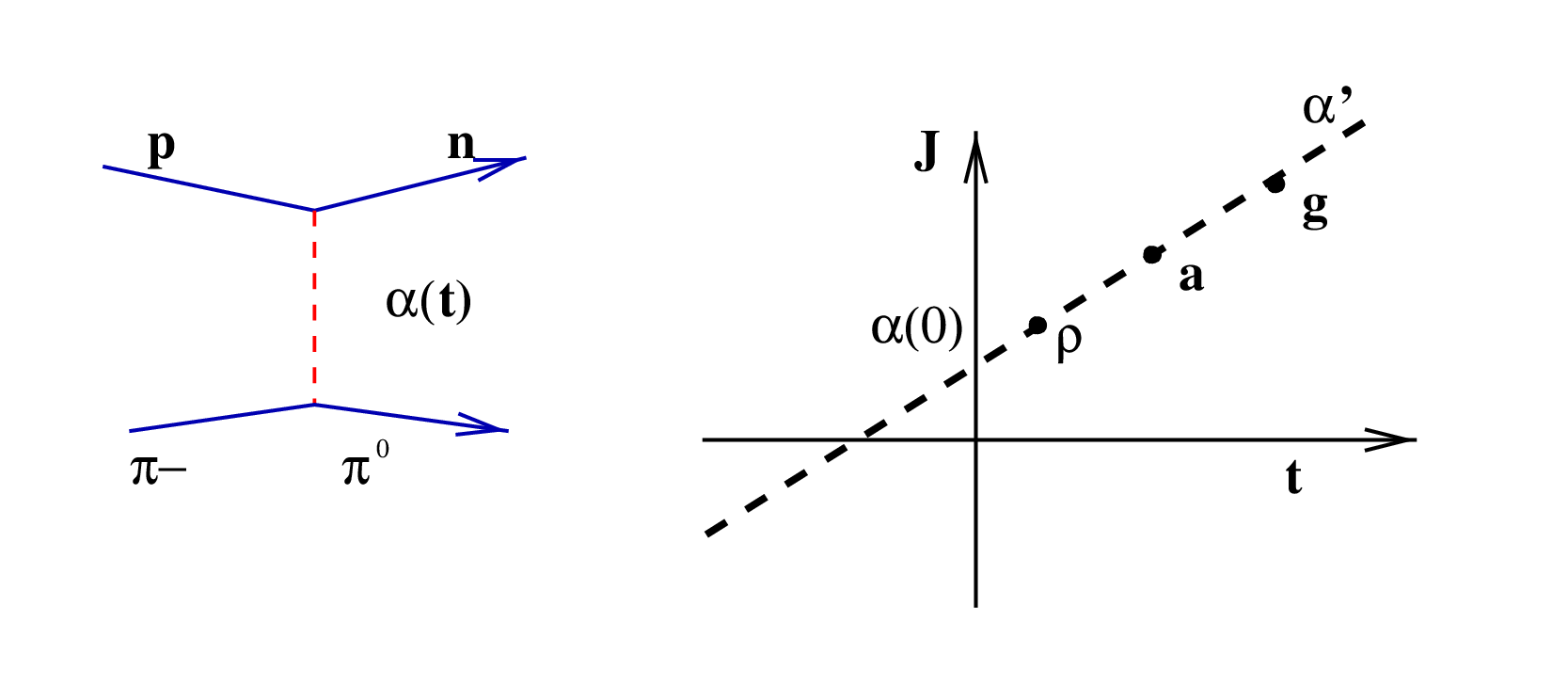}
}
\end{center}
\lm{{\protect\footnotesize  Schematic diagram for ${\pi^- p \rightarrow \pi^o n}$ scattering and the exchanged trajectory.}}
\label{fig:traj}
\end{figure}

\noindent Let's consider  the dependence of the  total cross section (a t-channel process)  with the square of the centre of mass energy  s. According to Regge theory it is parametrized as:
\be
{\rm \stot \propto \sum_k s^{\alpha_k(0)-1}}, \label{eq:stot1}
\ee

\noi where ${\rm \alpha_k(0) = 1,..n}$, are the intercepts of the  trajectories exchanged. Using only two main trajectories, ${\rm \stot}$  for ${\rm p\bar{p},\; pp, \; K^{\pm }p,\; \pi ^{\pm
}p,\; \gamma p}$ have been fitted by Donnachie and
Landshoff~\cite{ref:donlan} with an expression of the form: 

$${\rm \stot =  Xs^{0.0808}+ Ys^{-0.4525},}$$

\noi  where ${\rm X,\; Y}$ are parameters which depend on the exchanged field.  The first trajectory, called pomeron trajectory, has intercept  ${\rm \alphapom(0) = 1.0808}$  while
the second term, which represents an effective meson trajectory, has intercept ${\rm \alpha_k(0) = 0.545 }$. At high enough energy, only the pomeron term is important. The pomeron, identified as the first particle of the pomeron trajectory, is responsible for the rise of the total cross section as a function of the centre of mass energy.
 Since the  bulk of the  processes contributing to the total cross section  has very small ${\rm p_t}$, the pomeron  exchanged in these reactions is called `soft pomeron'. The soft pomeron trajectory has intercept  ${\rm \alphapom(0) \simeq 1.08 }$ and slope ${\rm \alpha^{\prime} \simeq 0.25}$ \gevtwom.


 Fig.~\ref{fig:dif_type} schematically shows  three  different types of diffractive reactions: elastic
scattering (a), single diffraction (b), where one of the incoming particle
dissociates, and double diffraction (c), where both incoming particles dissociate.
In diffractive scattering  the hadronization of the final states X and Y with masses ${\rm M_Y^2, \mxtwo}$ happens independently,
as shown in Fig.~\ref{fig:dif_type}(d). If the centre of mass energy ${\rm \sqrt{s}}$ is
large enough (ln(s) ${ \rm \gg ln(M_Y^2)+ ln(\mxtwo)}$), then
 there is a gap in rapidity between X and Y.
\bfm
\begin{center}
\leavevmode
\hbox{%
\epsfxsize = 4.in    
\epsffile{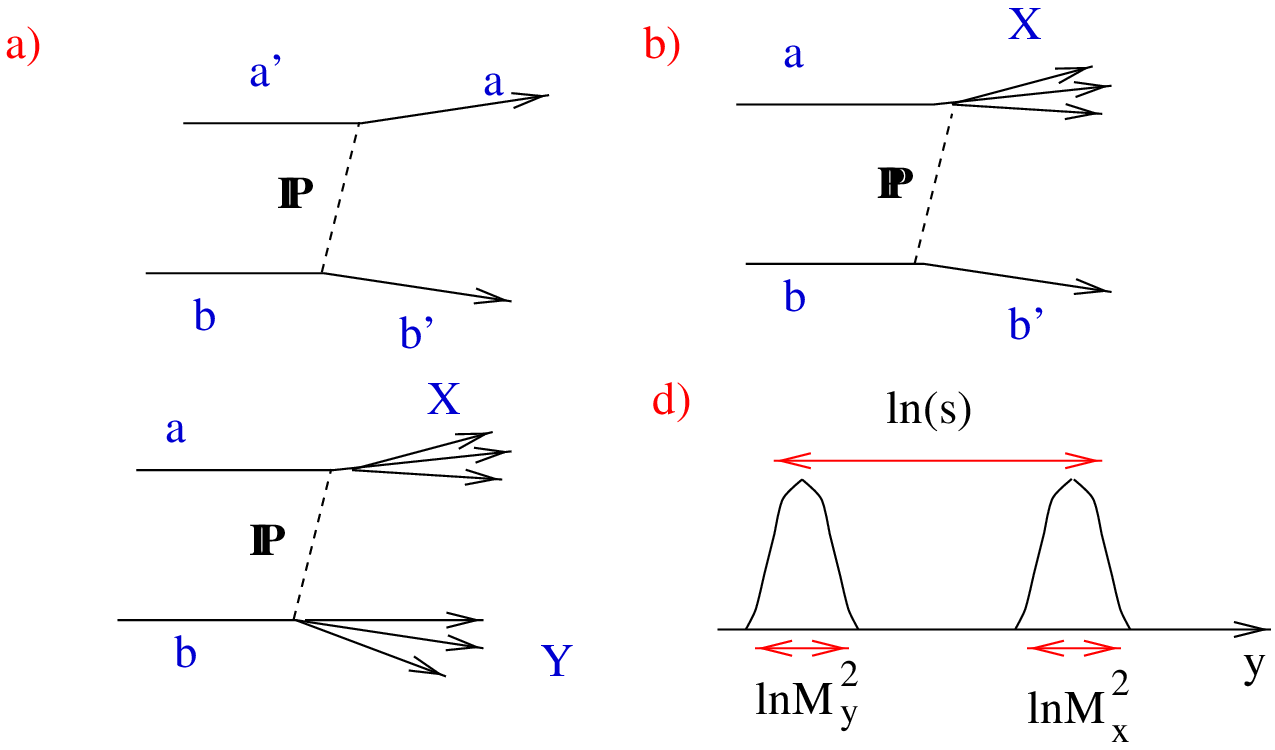}
}
\end{center}
\lm{{\protect\footnotesize Diagrams for three  different types of diffractive reactions: elastic
scattering (a), single diffraction (b) where one of the incoming particle
fragments and double diffraction (c). (d) shows energy flow as a function of rapidity for  ln(s) ${ \rm \gg ln(M_Y^2)+ ln(\mxtwo)}$.}}
\label{fig:dif_type}
\end{figure}

{


\section{Total  cross section at HERA}
The values of the total hadronic ${\rm \gamma p}$ cross section at HERA as measured by  the H1~\cite{ref:h1tot}  and ZEUS~\cite{ref:ztot}  collaborations  are shown in  Fig.~\ref{fig:sigmatot}  together with a compilation of low energy results. The Donnachie and Landshoff parametrizations including (dotted line) or not (solid line)
recent CDF~\cite{ref:cdftcs} results and  the  
ALLM~\cite{ref:hal} parametrization (dashed line) are also shown.
\begin{table}
\begin{center} \begin{tabular}{|c|c|c|} \hline
 Experiment & \Wgpm range & \Wgpm  \\ \hline
ZEUS &  $167 < \Wgpm < 194$ &  $\stgp = 143 \pm 4 \pm 17$ $\mu$b  \\  \hline
H1 & $<\Wgpm>$=200 &  $\stgp  =165 \pm 2 \pm 11$ $\mu$b  \\  \hline
\end{tabular}\end{center} 
\caption{\protect\footnotesize Summary of experimental results on the measurements of ${\rm \sigma_{tot}}$ at HERA.} 
\label{tab:stot}
\end{table} 
The HERA data are in agreement with these predictions and therefore with  the assumption that  also at HERA  `soft' pomeron exchange is responsible for the increase  of \stgp  as a function of  the centre of mass  energy. 

The diffractive cross section represents a large fraction of the total cross section: at HERA, for example, 
the diffractive and non diffractive parts are, according to the H1 collaboration~\cite{ref:h1tot} ,  ${\rm \sigma _{dif}
^{\gamma p} = 69.2 \pm 13.2}$ $\mu$b and ${\rm \sigma _{non-dif} ^{\gamma p} = 96.1
\pm 17.9}$ $\mu$b, giving ${\rm \sigma_{dif}/\sigma_{tot} = (42\pm 8) \%  }$, while according to the ZEUS collaboration ~\cite{ref:ztot} \nl 
${\rm \sigma_{dif}/\sigma_{tot} = (36\pm 8) \%  }$.


\bfm
\begin{center}
\leavevmode
\hbox{%
\epsfxsize = 4.in
\epsffile{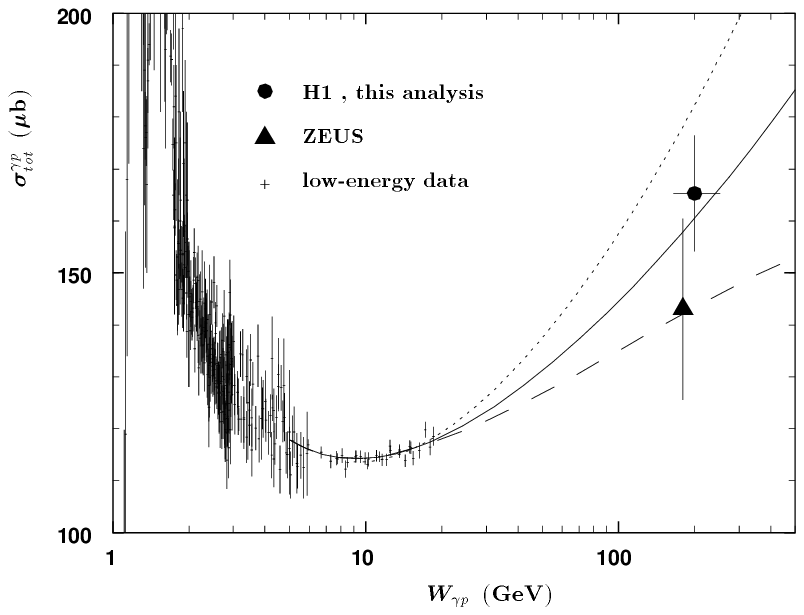}}
\end{center}
\lm{{\protect\footnotesize Total ${\rm \gamma p}$ cross section as a function of \Wgpm. The results are shown together with two parametrizations from Donnachie and Landshoff  that  include (dotted line) or not (solid line) recent CDF  results and with the ALLM parametrization  (dashed line).}}
\label{fig:sigmatot}
\end{figure}

\section{Kinematics of diffractive events at HERA}

In  Fig.~\ref{fig:diffr}, a diagram for diffractive ep scattering is shown.
A photon ${\rm \gamma^*(Q^2)}$ with virtuality ${\rm -q^2 =}$\qtwom is emitted at the electron vertex\footnote{ The symbol ${\rm \gamma }$ is used for quasi real photon while the symbol ${\rm \gamma^*}$ is used for virtual photon.} . Depending on the value of \qtwom, the events are divided into two large families: photoproduction, for ${\rm Q^2 < 4}$ \gevtwom,  $\;$ and deep inelastic scattering (DIS), for ${\rm Q^2 >4}$ \gevtwom.
 ${\rm s=(k+p)^2}$ is defined as the centre of mass energy squared  of the ep system  while    ${\rm \Wgpm^2=(q+p)^2}$ is used to indicate  the centre of mass energy of the virtual photon-proton (\gpm) system.
At large \qtwom, in  the frame where the proton has infinite momentum, the variable \xbj = ${\rm \frac{Q^2}{2 p \cdot q}}$ represents the fraction of the proton longitudinal momentum carried by the struck quark.  In the proton rest frame,  ${\rm y=Q^2/(s\xbj)}$  equals  the fraction of the electron energy transferred to the proton.

In addition, diffractive events
 are described by the following variables: t, the square of the
four-momentum transfer at the proton  vertex and \xpom, the momentum fraction of the pomeron in the proton.

\bfm
\begin{center}
\leavevmode
\hbox{%
\epsfxsize = 4.in
\epsffile{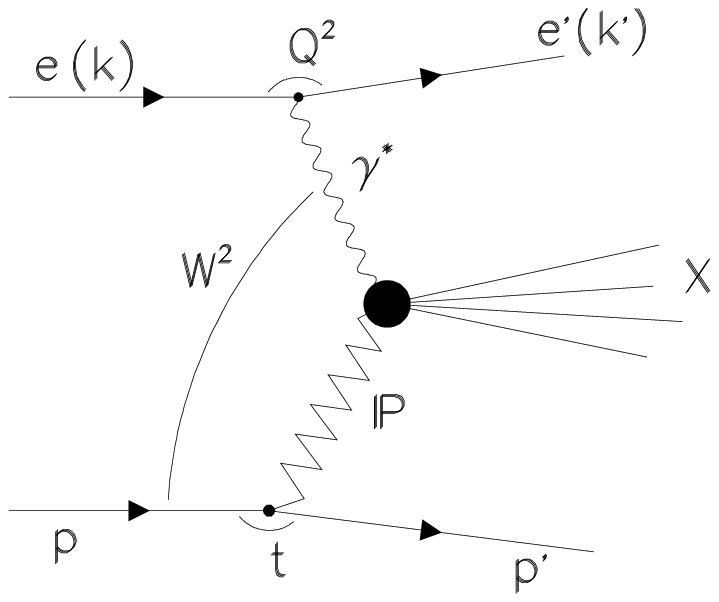}}
\end{center}
\lm{{\protect\footnotesize Diagram for diffractive scattering at HERA. }}
\label{fig:diffr}
\end{figure}
\noi If the reaction is elastic or single diffractive (or photon diffraction as
sometimes
 single diffraction is called for the HERA regime), then the quantities t
and \xpom  can be determined either from the scattered proton or from the 
system \mx.

If the longitudinal and transverse
momentum  of the  scattered proton, ${\rm p^{\prime}_z,p^{\prime}_\perp}$,  are measured, then
\xpom and t are calculated as: 

\be 
{\rm  x_L \simeq p^{\prime}_z/E_p \longrightarrow \xpom = 1- x_L}
\ee

\begin{equation}
{\rm
t=(P-P^{\prime})^2\simeq -\frac{ (p^\prime_\perp)^2}{x_L} - m_p^2 \frac{(1-x_L)^2}{x_L} }
\end{equation} where ${\rm m_p}$ is the proton mass.
If the proton is not observed, a measurement of \xpom  can be obtained as:
\begin{equation}{\rm  \xpom = \frac{(P-P^{\prime})\cdot q}{P \cdot q} \simeq
        \frac{\mxtwo+Q^2}{W^2+Q^2-m_p^2} }, \end{equation}

\noi where \mx is the mass of the system X. t can be reconstructed from the
system X only for some exclusive reactions, such as vector meson production, 
 where the resolution on ${\rm p^\prime_\perp}$ is accurate enough.



\section{Experimental signature of diffraction at HERA}
\label{sec:expsig}
One of the main issues concerning diffraction at HERA is the experimental method to
separate diffractive from non-diffractive events. For some exclusive reactions
the distinction is actually quite easy. Let's consider  for example exclusive $\rho^0$ production and decay: 
$${\rm \gamma p \longrightarrow \rho^0 p}$$
$${\rm \rho^0 \longrightarrow \pi^{+}\pi^{-}.}$$
The central detector is empty,  except for the two tracks coming
from the $\rho^0$ decay. This topology is very unusual and the background from `non-pomeron' exchange is negligible. Inclusive ${\rm \gamma}$ diffraction, ${\rm \gamma p \ra X p}$, is on the other hand more difficult to identify. 
Two quantities can help in the distinction: a rapidity gap in the
final state particles production and/or  the presence of a highly energetic
scattered  proton.

\subsection{Rapidity gaps}
\bfm
\begin{center}
\leavevmode
\hbox{%
\epsfxsize = 3in
\epsffile{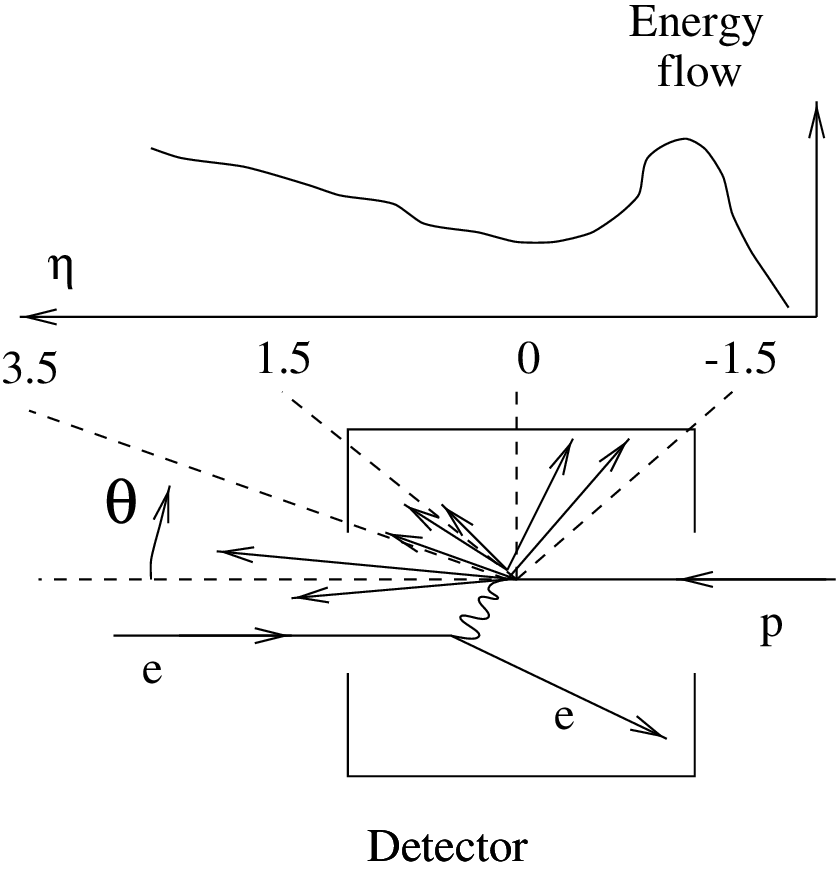}}
\end{center}
\lm{{\protect\footnotesize Energy flow vs $\eta$ for non diffractive ep scattering at HERA. }}
\label{fig:rap_gap_dis}
\end{figure}

\noi Fig.~\protect\ref{fig:rap_gap_dis} schematically shows the energy flow as a function of pseudorapidity ${\rm \eta}$ for non diffractive ep scattering at HERA\footnote{ The pseudorapidity $\eta$ is defined as:  ${\rm \eta=-ln(tan( \theta/2))}$. Following the  HERA convention,  the angle $\theta$ is measured with respect  of the proton beam direction. } .
Aside from the recoil electron, two main groups of particles can be identified: particles produced at high rapidity in the hadronization of the proton remnant, and particles produced in the hadronization of the photon-parton system, typically at small or negative rapidity. In deep inelastic scattering, for example, the struck parton is
deflected and emerges from the proton remnant at
an angle ${\rm \theta_{q}} $. It is useful to express this angle as the difference  in  pseudorapidity 
 between the  struck parton and the proton remnant: 
\begin{equation} {\rm
\Delta\eta = \eta _{ proton \; remnant} - \eta _{ \; parton} }. 
\end{equation} 
 \noindent Since the pseudorapidity
interval covered by a system with centre of mass energy \rs $\;$ is given by:

\begin{equation}{\rm \Delta\eta \sim ln(\frac{s}{m_p^2}) }
\end{equation} 
with
m$_p$ the proton mass, then we can show that  the pseudorapidity interval 
between the proton remnant  and the struck quark is:  
\begin{equation} {\rm
\Delta\eta  \sim ln(\frac{\Wgpm^2}{m^2_p}) - ln(\frac{x_{Bj}\Wgpm^2}{m^2_p}) \sim
ln(\frac{1}{x_{Bj}}),}  
\end{equation} 
where ${\rm ln(\frac{W_{\gamma p}x^2}{m^2_p})}$ is the
total rapidity covered by the $\gamma$-p system and  ${\rm
ln(\frac{x_{Bj}W_{\gamma p}^2}{m^2_p})}$ is the amount covered by the $\gamma$ - struck quark system.
\\ Due to the  colour  string connecting the struck parton and the proton
remnant  the rapidity gap ${\rm \Delta\eta}$   is filled with particles
in  the hadronization process. In particular  as \xbj decreases, the  average
hadron multiplicity ${\rm <n_{h}>}$   increases faster than  the pseudorapidity
interval ${\rm \Delta\eta}$  making it less and
 less likely  for rapidity gaps to be visible~\cite{ref:ryb} . If we assume the produced
hadrons to fill the rapidity gap according to  a Poisson distribution,  the probability
${\rm w_{gap}}$ to have no particles in the gap ${\rm \Delta\eta}$ has the
form:  
\begin{equation}
 {\rm w_{gap} \sim e^{- <n_{h}>} <  e^{- \Delta\eta} . } \label{eq:exp}
\end{equation}
 This expression means that   rapidity gaps between   the
proton fragments and the jet produced by  the struck quark  are exponentially
suppressed. 

\bfm
\begin{center}
\leavevmode
\hbox{%
\epsfxsize = 3in
\epsffile{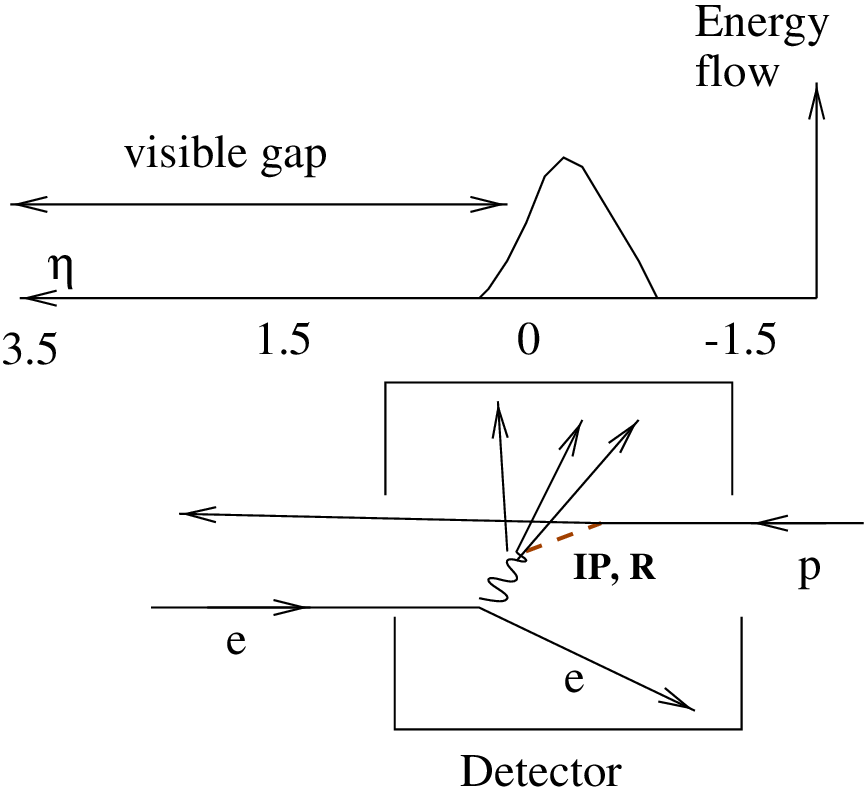}}
\end{center}
\lm{{\protect\footnotesize Energy flow vs $\eta$ for  diffractive ep scattering at HERA. }}
\label{fig:rap_gap_dif}
\end{figure}
\noi For  Reggeon or Pomeron exchange, Fig.~\protect\ref{fig:rap_gap_dif}, the probability to have a 
rapidity  gap $\Delta\eta$ depends on the intercept of the exchanged trajectory\cite{ref:kop}:
$${\rm p(\Delta\eta) \sim e^{-2 ( \alpha(0) -1)\Delta\eta} .}$$
Let's then consider different possibilities:
\bi
\item [] \pomer exchange: ${\rm \alpha_{\pom}(0) \sim 1 \;\; \Rightarrow 
\;\; p(\Delta\eta) \sim e^{0}}$ 
\item  []${\rm \rho,a_2, f_2,\omega}$ exchange: ${\rm \alpha_R(0) \sim 0.5 \; \;
\Rightarrow \;\; p(\Delta\eta) \sim e^{-\Delta\eta}} $ 
\item []${\rm \pi}$ exchange: ${\rm \alpha_{\pi}(0) \sim 0 \; \;
\Rightarrow \;\; p(\Delta\eta) \sim e^{-2\Delta\eta}}. $ 
\ei 

\noi Therefore, even though  ${\rm \rho, \pi \; and \; \pomer}$ are colourless exchanges,
only \pomer exchange produces rapidity gaps that are not suppressed as the gap width increases. It is therefore possible to operationally define  diffraction~\cite{ref:bj}
by the presence of a rapidity gap: diffractive events are those which lead to a large
rapidity gap in final state phase space and are not exponentially suppressed as a function of the gap width.

\subsection{Leading proton in the final state}

In diffractive events, the incoming beam particles, when they do not dissociate, conserve a large
fraction of their initial momentum. At HERA the diffractively scattered proton carries on
average more than 99\% of its initial momentum. The cross sections for non diffractive processes to produce so energetic  protons is very small compared with the diffractive cross section making  the detection of a high energy proton a clean tag for diffractive physics. Fig.~\ref{fig:xl} schematically shows the spectra
of leading protons generated from different mechanisms: at \xl $\simeq 1$
single diffraction is almost the sole component, while moving away
from \xl = 1 double diffraction and reggeon exchange become important. Traditionally, \xl=0.9 has been used to indicate the \xl value at which the diffractive and not diffractive part of the  spectrum are equal. 
Leading protons can also be produced in  `standard' DIS events as  part of the 
proton remnant jet, but they have on average  a much lower \xl value. A  recent release of the LEPTO MonteCarlo~\cite{ref:ing} , on the other hand, includes leading protons production in the fragmentation of the proton remnant with a cross section comparable to reggeon exchange. 
Note that the distinction between the different mechanisms for leading protons production is somehow arbitrary
 and there might be  a lot of overlap.

\bfm
\begin{center}
\leavevmode
\hbox{%
\epsfxsize = 4in
\epsffile{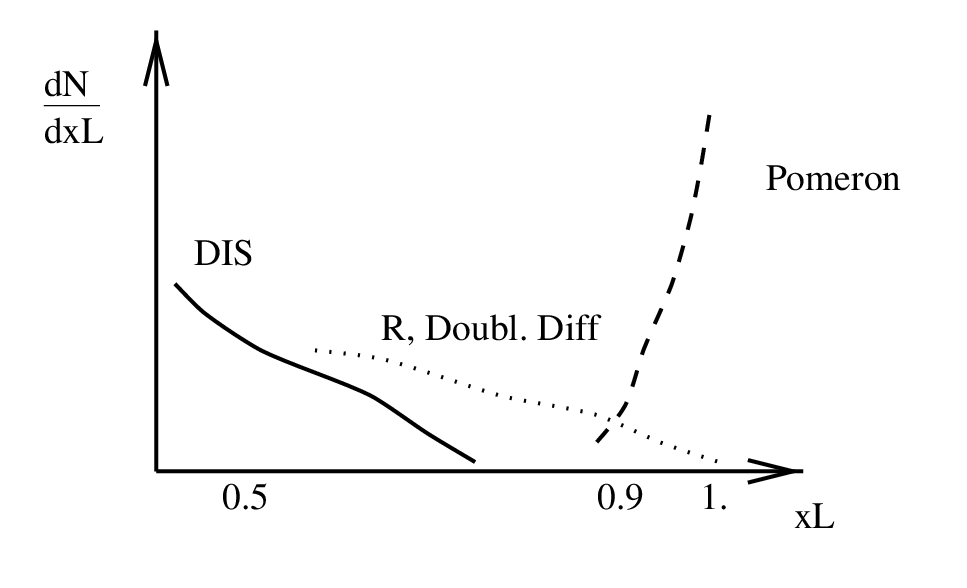}}
\end{center}
\lm{{\protect\footnotesize  Spectra
of leading proton generated from different mechanisms: pomeron exchange (dashed line),
reggeon exchange and double diffraction (dotted line) and  `standardÕ DIS (solid line) as a function of \xl.
}}
\label{fig:xl}
\end{figure}
\noi In the transverse plane, leading protons have rather small momentum, with a typical ${\rm  p_\perp^2}$ distribution of the form:

$${\rm\frac{dN}{dp_\perp^2} \sim e^{-b\cdot p_\perp^2}} $$
 with b =  5-15 ${\rm GeV^{-2}}.$

\noi This feature makes their detection quite difficult since they
tend to stay very close to the beam line. Because of this, movable sections of the beam pipe,
called `Roman pots', are used by  both the H1 and ZEUS
collaborations to allow the insertion of high precision detectors down
to few centimeters  from the beam line.

\section{Models for diffractive ${\rm \gamma^{*} p}$ scattering}
\label{sec:model}


Several models have been proposed to explain diffractive interactions  in ep scattering. In some instances, a connection is made between Regge concepts (like Pomeron) with QCD concepts (like gluons). Here we present a brief  description of some of  the ideas on which the models are based.

\par
- {\bf Factorization of vertices and pomeron structure function:} Factorization  considers 
 the ${\rm {I\! P} p}$ vertex as  independent of  the  ${\rm {I\! P} \gamma^{*}}$ interaction. 
A universal pomeron flux factor  ${\rm f_{I\!P/p}(x_{I\!P}, t)}$ characterises
 the  ${\rm {I\! P} p}$ vertex and parametrizations 
obtained from  fits to pp and  p\={p}  diffractive data can be  used in ep collisions.
 Several different expressions for $\;{\rm f_{I\!P/p}(x_{I\!P}, t)}$ have been
proposed\cite{don84,don87,ingsch} which all include an exponential dependence on t of the type ${\rm e^{-b\mid t \mid}}$ with b$\; \sim $ 5-8 \gevmtwom $\;$ and a dependence on the pomeron longitudinal momentum \xpom of the type ${\rm \sim1/x_{I\!P}}$ . 
These models express the diffractive 
  $\gamma^*$ p cross section as
\begin{equation}{\rm
\sigma^{\gamma^* p}(Q^2,\beta,\xpom,t) \; \propto  f_{I\!P/p(x_{I\!P}, t)} \cdot \sigma^{\gamma^{*} {I\!P}}(Q^2,\beta),
}\end{equation}
\par
\noi and  describe deep inelastic $\gamma^{*}$p and 
 ${\rm \gamma^{*} {I\!P}}$ interactions  in the same way:
 the incoming  $\gamma^{*}$ interacts with
one component of the target  leaving behind a remnant.
 For ${\rm \gamma^{*} {I\!P}}$ interaction the scaling variable  
that plays the same role of  \xbj  is:
\begin{equation}{\rm
\beta = \frac{Q^2}{2x_{I\!P}p_{proton}\cdot q} \sim \frac{Q^2}{\mxtwo +Q^2} }.
\end{equation} 
\noi The ${\rm \gamma^{*} {I\!P}}$ cross section can then be written as:
\begin{equation}{\rm
\sigma_{tot}^{\gamma^{*} {I\!P}}(Q^2, \beta) = 
\frac{4\pi^2\alpha_{em}}{Q^2} F_2^{I\!P}(Q^2, \beta),
}\end{equation}
where  F${\rm _2^{I\!P}(Q^2, \beta)}$ is the  pomeron structure 
function; ie, the probability for finding a quark of fractional momentum $\beta$ in the pomeron. \par

- {\bf Factorization breaking  and effective pomeron structure function:}
The diffractive  ep interaction is viewed as photon diffractive dissociation 
on the proton (some examples are given in \cite{nik91,abr,ref:nz2,gots,levin}). Consider  ep scattering in the proton 
rest frame. Upstream  of the proton, Fig.~\ref{fig:lc}, the incoming $\gamma$ (or ${\rm \gamma^*}$) 
fluctuates  into different hadronic states and  its wave function can be expressed as: $${\rm |\gamma>=|\gamma>_{bare}+|q
\bar{q}> +|q \bar{q}g>+.. .}$$  

\noi The pomeron, viewed as two gluon-exchange, couples to these hadronic states  in an s-channel interaction
and, technically, there is no \pom $\;$ remnant since both gluons interact with the photon. In this approach the 
virtual photon couples with more than one pomeron constituent  and the meanings of $\beta$
and  F${\rm _2^{I\!P}(Q^2, \beta)}$ are not well defined in terms of partons; in particular, factorization is not a natural consequence. To compare the predictions from this approach
 to those of the previous type of model, an effective \pom $\;$ structure function,
 F${\rm _{2 \; eff.}^{I\!P}( Q^2, \beta)}$,  is introduced.
In this picture, the interaction of the ${\rm q\bar{q}}$ state generates a different value of $\alphapom$ than the interaction of the ${\rm q\bar{q}g}$ state~\cite{genovese} with  an effective $\alphapom$ increasing at small \xpom (\xpom $\simeq\; 10^{-4} $). Therefore the  ${\rm q\bar{q}}$ and  ${\rm q\bar{q}g}$ fluctuations have different pomeron fluxes breaking the factorization mechanism. According to~\cite{ref:nz3} , factorization is also broken by the exchange of longitudinal photons. The states are characterised by the transverse  and longitudinal momentum (${\rm k_\perp, z}$) of the \qqbar pair (taken as an example) and by the quark mass (${\rm m_q}$).  The radius ${\rm r^2_\perp}$ of the state depends on the inverse of ${\rm  k_\perp^2\cdot Q^2}$ and ${\rm m^2_q }$:
$${\rm \frac{1}{r_\perp^2} \propto \frac{k_\perp^2\cdot Q^2}{\mxtwo}}$$
$${\rm \frac{1}{r_\perp^2} \propto   m^2_q }.$$

\bfm
\begin{center}
\leavevmode
\hbox{%
\epsfxsize = 4in
\epsffile{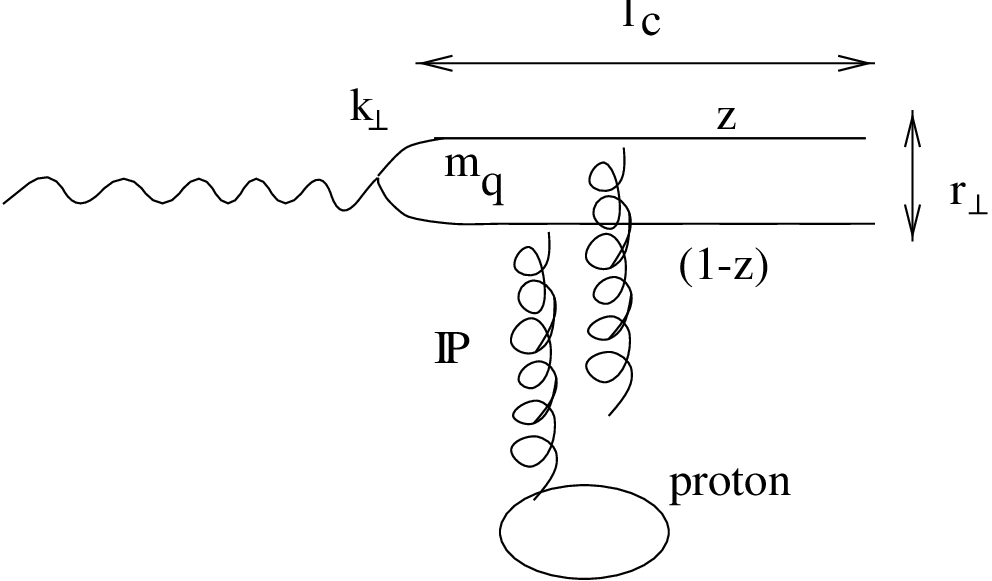}}
\end{center}
\caption{{\protect\footnotesize Diffractive ${\rm \gamma p}$ scattering in the proton rest frame.}}
\label{fig:lc}
\end{figure}
\noi If the state has large ${\rm r^2_\perp}$, both ${\rm  k_\perp^2\cdot Q^2}$ and ${\rm m^2_q }$ are small. In this case, the gluon-quark coupling is large, and pQCD cannot be applied since the  photon acts like an extended object providing no information on the microscopic nature of the interaction.  Large size fluctuations are thought to be responsible for the rise of the total hadronic cross section with energy and represent the bulk of diffractive events. Large ${\rm r^2_\perp}$
configuration have small  ${\rm  k_\perp}$ and the final state particles tend to be aligned along the photon-pomeron axis. This phenomenology is know as the Aligned Jet Model~\cite{ref:alj} .

 Conversely, if ${\rm r^2_\perp}$ is small either
because  ${\rm  k_\perp^2\cdot Q^2}$ (high \qtwom DIS events) or ${\rm m^2_q }$ (production of charm or bottom \qqbar pair) is large, then the  gluon-quark coupling is small and pQCD can be applied.

\par


\par

- {\bf Soft color interaction:} in this model~\cite{ref:buch} , diffractive scattering  is viewed as dominated by the exchange of one `hard' gluon plus non-perturbative color interactions to allow a color singlet final state.

\section{Vector meson production: ${\rm \gamma p}$ \ram Vp} 

\noi In the range of centre of mass energy \Wgpm up to  20 GeV covered by pre-HERA experiments, this process has been described very successfully within the framework of Vector Dominance Model (VDM)\cite{ref:fey} . In this model, the photon is assumed to fluctuate into a virtual vector meson which then interacts elastically with the proton via the exchange of a pomeron, Fig.~\ref{fig:vm}(a).
 \noi From VDM one expects:
\begin{equation}{\rm
\sigma^{\gamma p \rightarrow V p} = \frac{4 \pi \alpha }{f^{2}_{V}} \sigma^{Vp \rightarrow Vp},}
\label{eq:vdm}
\end{equation}
where  ${\rm f_V^{2}/4\pi}$ is the photon-vector meson coupling constant, which expresses the fact that the ${\rm \gamma p}$ cross section should behave as an hadronic cross section. 

\noi Since vector meson production represents the elastic part of \stgp, we can use  the optical theorem to  relate the two cross sections at t = 0 \gevtwom $\;$:

\be {\rm
(\frac{d\sigma^{\gamma p \rightarrow V p}}{dt})_{t=0} = A\cdot \sigma_{tot}^2,
}
\ee
\noi with A=constant. Then  we can express the elastic cross section at any t value as:
\be {\rm
\frac{d\sigma^{\gamma p \rightarrow V p}}{dt}(t) =  (\frac{d\sigma^{\gamma p \rightarrow V p}}{dt})_{t=0}\cdot e^{f(t)}= A\cdot \sigma_{tot}^2\cdot e^{f(t)},
\label{eq:sigtot}
}
\ee
\noi where ${\rm f(t)}$ is the functional dependence of the cross section on t. For vector meson production, according to Regge theory, f(t)  can be written as:
\be {\rm f(t) =t\cdot( b_0 + 2\alphapom^\prime \cdot ln(\Wgpm^2/W^2_0))},
\label{eq:ft}
\ee 
\noi where ${\rm b_0 \;\; and \;\; W^2_0}$ are parameters. Using eq.\ref{eq:ft} into eq.\ref{eq:sigtot}, integrating over t and writing explicitly the dependence from the centre of mass energy \Wgpm, we obtain:
\be {\rm
 \sigma^{\gamma p \rightarrow V p}(\Wgpm) \propto \frac{ (\Wgpm^{2})^{2 \cdot(\alpha _{\pom}(0) - 1)} }{ b_0 + 2\alphapom^\prime \cdot ln(\Wgpm^2/W^2_0)} \simeq  \Wgpm^{0.22}
.}
\label{eq:wdep}
\ee 

\noi It is very important to note that this reasoning is based on the assumption, supported by pre-HERA data, that the same underlying exchange governs both the total and vector meson cross section.

Recent papers~\cite{ref:ry,ref:br,ref:ne} have shown that the situation at HERA, given the large value of \Wgpm, might be qualitatively different if a hard scale is present in the interaction. Under these  circumstances, the process is calculable in QCD. The approach outlined in Section (\ref{sec:model}) is used:  the photon fluctuates into a \qqbar pair which first interacts with the target and then  the meson is formed~Fig.~\ref{fig:vm}(b). 
\bfm
\begin{center}
\leavevmode
\hbox{%
\epsfxsize = 4in
\epsffile{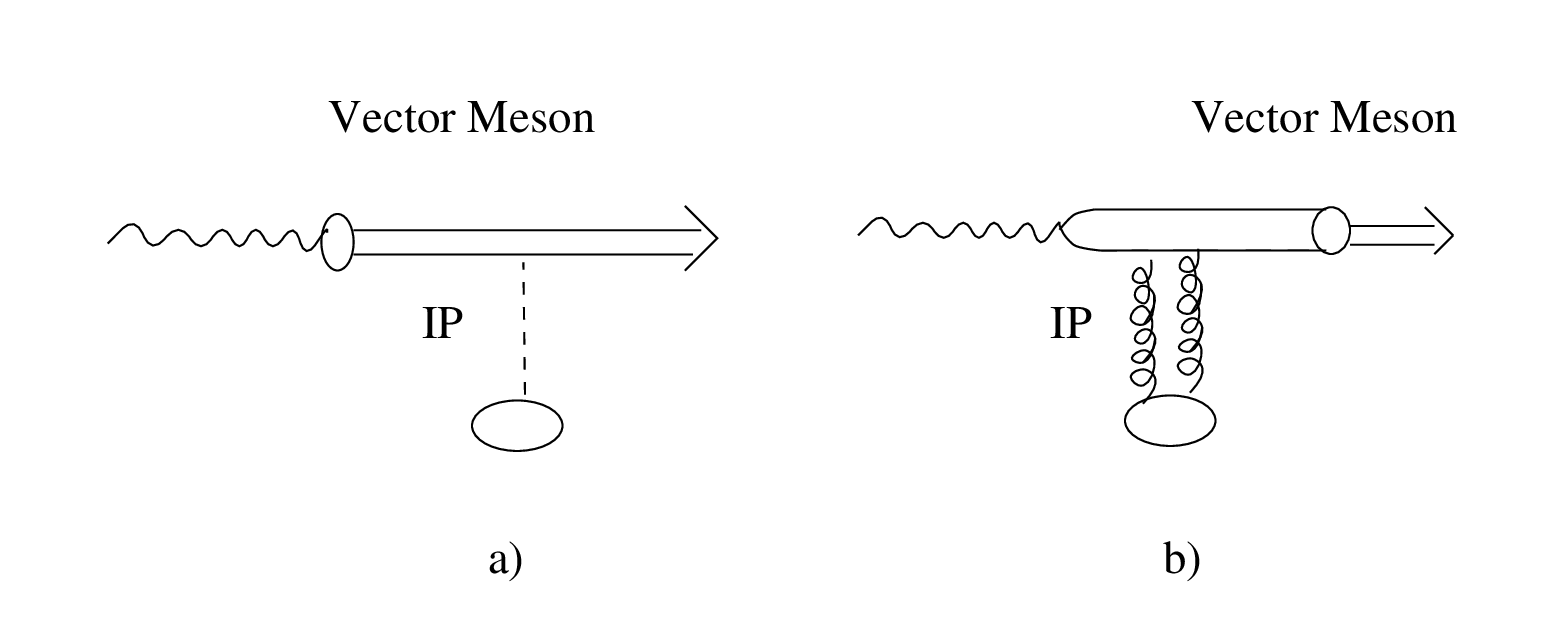}}
\end{center}
\lm{{\protect\footnotesize Different models for ${\rm \gamma p \ra V p}$: (a) vector meson dominance, (b) photon diffraction.}}
\label{fig:vm}
\end{figure}
The scale of the interaction is given by the reciprocal of the fluctuation radius  ${\rm r^2_\perp}$ and therefore, if either ${\rm k_\perp^2\cdot Q^2}$, t or ${\rm m^2_q}$ is large, the process 
is hard.
Since the  transverse momentum ${\rm  k_\perp^2}$ generated at the photon-quark vertex is different for
longitudinally and transversely polarised photons, with ${\rm \sigma^{\gamma_L \cdot p}}$ dominated by large  ${\rm k_\perp^2}$ and ${\rm \sigma^{\gamma_\perp  \cdot p}}$   dominated by small  ${\rm k_\perp^2}$,
 early papers dealt only with the former photon polarisation. Lately~\cite{ref:ry2} also predictions for ${\rm \sigma^{\gamma_\perp  \cdot p}}$ have been made.

The pQCD approach has been used to calculate the magnitude and energy dependence of the cross section  for  photoproduction of $J/\psi$ mesons~\cite{ref:ry} , where the charm mass ensures a hard scale, and production of $\rho^0$ mesons at high \qtwom~\cite{ref:br} . In both cases, the \qqbar pair resolves the gluonic
contents of the proton giving a cross section proportional to the proton gluon distribution squared\footnote{The square comes trivially from the fact that the pomeron in made of two gluons}:
\begin{equation}{\rm
 \sigma^{Vp} \propto 
[\alpha_s (\bar q^2) \bar x g(\bar x,\bar q^2)]^2. 
\label{eq:2g}
}\end{equation}
\noi The energy dependence is therefore no longer determined  by the pomeron intercept but by the rise of the gluon distribution at low \xbj.

\subsection{Experimental signature and selection methods for ${\rm \gamma p \ra V p}$}
Vector meson production is characterised by very little activity in the detector since only the vector meson decay products and, for the DIS case, a scattered electron, are present.
The  processes studied so far by the two collaborations are\footnote{References from \cite{ref:zw1} to  \cite{ref:rhop1} are contribution to ICHEP 1996, Warsaw} listed in Tab.2 .
\begin{table}
\bc \begin{tabular}{| l | c | c | c| c|}\hline
VM decay mode & \qtwom $\simeq$ 0 $\;$ \gevtwom & ref. &\qtwom $>$ 0 $\;$ \gevtwom   & ref. \\ \hline
 ${\rho^0 \longrightarrow \pi^+ \; \pi^-}$ &X&\cite{ref:zrho2,ref:hrho1,ref:zrho3,ref:zw1} &X  &\cite{ref:zrho1,ref:hrhoj1,ref:zw2,ref:zw3}\\ 
 ${\phi \longrightarrow k^+ \; k^-}$ &X&\cite{ref:zphi1} &X &\cite{ref:zphi2,ref:zw3,ref:hw2}\\
 ${J/\Psi \longrightarrow \mu^+ \; \mu^-, \;\;e^+e^-}$ &X&\cite{ref:hjp1,ref:zjp1,ref:hjp2} &X &\cite{ref:hrhoj1,ref:zw3}\\ 
 ${\omega \longrightarrow \pi^+ \; \pi^- \; \pi^0}$ &X&\cite{ref:zw4} & & \\
 ${\Psi(2S) \longrightarrow \pi^+ \; \pi^- \;J/\Psi }$ &X&\cite{ref:jp2s1} & &\\
 ${\rho^{'} \longrightarrow \pi^+ \; \pi^- \;\pi^+ \; \pi^-  }$ && &X &\cite{ref:rhop1}\\ \hline
\end{tabular}
\caption{\pf List of decay modes used to identify vector meson production  ${\rm \gamma p \ra V p}$ at HERA.}
\end{center}
\protect\label{tab:mode}
\end{table}
\nl
\noi General requirements common to the analyses dealing with vector meson production are:
\bi
\item [-] Predictable number of tracks for a given reaction,
\item [-] Energy clusters in the calorimeter matching the tracks momenta, with a maximum unmatched energy of ${\rm \sim 0.5 - 1}$ GeV (determined by resolution),
\item [-]{ \Wgpm range restricted to 40-140 GeV. For small (large) \Wgpm  values, the tracks are too forward (backward) to be measured in the tracking chamber. For some analyses, higher values of \Wgpm have been achieved using events with the vertex displaced in the forward direction and/or using the calorimeter.}
\ei

\noi The main sources of systematic errors 
come  from uncertainty on the trigger thresholds, input MonteCarlo distributions, track reconstruction, uncertainties in the mass fit (in particular for 
the $\rho^{0}$ analysis), non resonant background subtraction,  and magnitude of the double dissociation contribution.
This last contribution is of particular interest since very little is known about double diffractive production of vector mesons. Fig.~\ref{fig:dd} shows the contamination mechanism: if the mass  ${\rm M_Y}$ of the excited proton system is small  ($\le $1.6  GeV  for H1, $\le $ 2 GeV for ZEUS depending on each detector forward coverage), the event looks  elastic  and  is included in the sample. On the contrary, if ${\rm M_Y}$ is large, energy deposition in the calorimeter can be used as a veto.  The CDF Collaboration~\cite{ref:cdfdd} measured the mass spectrum of the system Y in ${\rm p\bar{p}}$ diffraction to be ${\rm dN/dM_Y^2 = 1/M_Y^n}$ with n = 2.2 . 
This result, however, has not been obtained in the very low mass region and therefore should be used only as an indication.
A more direct method used by both the H1 and ZEUS collaborations  is to
model the visible energy deposition due to high  ${\rm M_Y}$ states as a function of n to determine which value fits  the data best and use it to perform the background subtraction.

\bfm
\begin{center}
\leavevmode
\hbox{%
\epsfxsize = 4.in
\epsffile{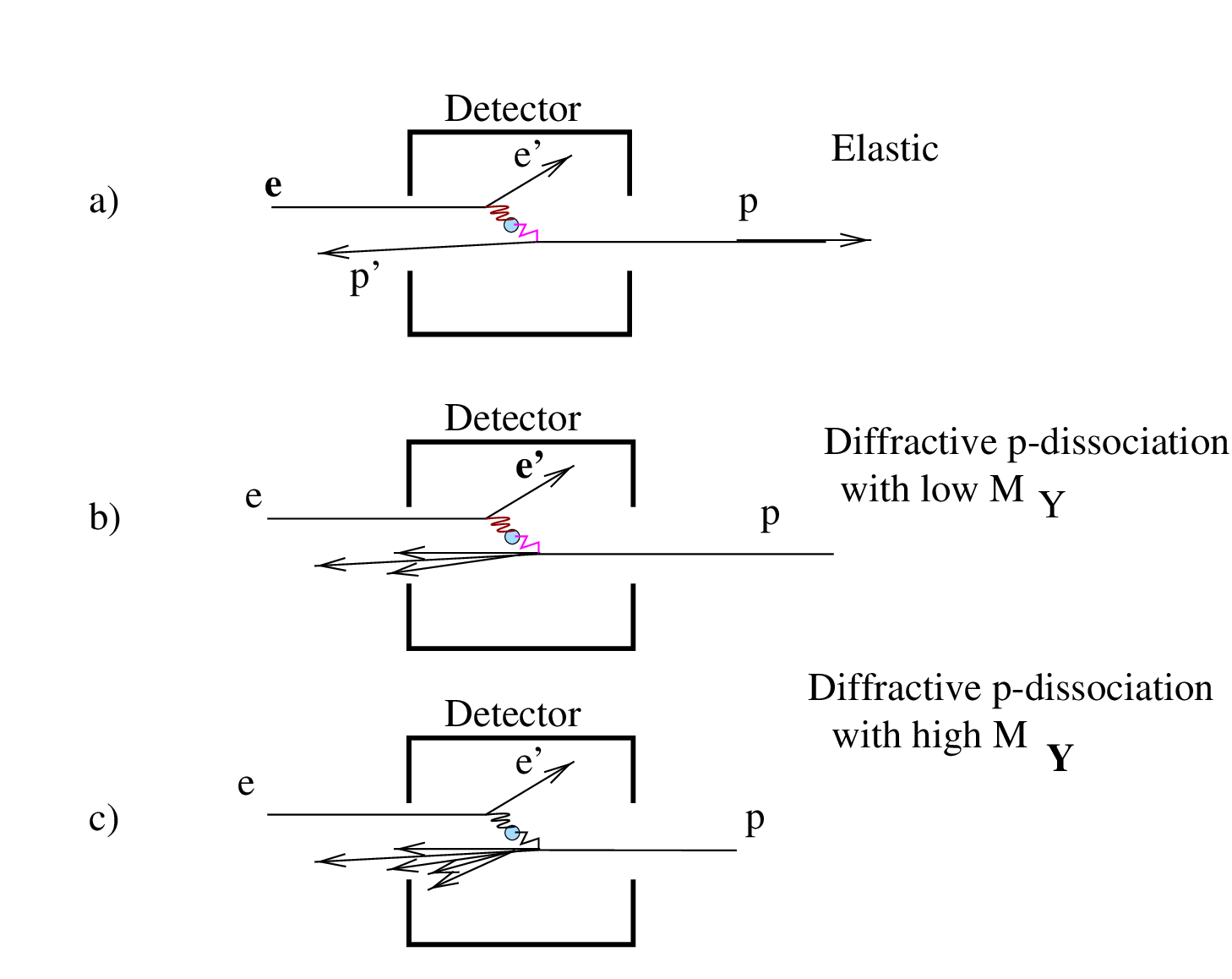}}
\end{center}
\lm{{\protect\footnotesize Double dissociation background to single dissociation.  b) If the mass $M_Y$ of the excited proton system is small  the event looks like single diffraction and  is included in the sample of elastic vector meson production. c) For large $M_Y$, energy deposition in the calorimeter can be used as a veto.}}
\label{fig:dd}
\end{figure}

\subsection{Light vector meson production at \qtwom = 0}
The energy dependence of the total cross sections for light mesons ($\rho^0$, $\omega$, $\phi$) and J/$\psi$ photoproduction, as measured 
by ZEUS and H1 are shown in Fig.~\ref{fig:vm_photo_cs}. The Regge theory prediction ${\rm \sigma^{\gamma p\ra Vp}(W_{\gamma p})\propto W_{\gamma p}^{0.22}}$ is clearly supported by the data. 
 A summary of experimental results on the measurements of ${\rm \sigma_{tot}, \; b \; and\; r_{00}^{04}}$ in photoproduction is given in Tab.~\ref{tab:bphp}.
 The b values are consistent with parametrizations of low energy data and with the `shrinkage mechanism' expected in Regge theory. This mechanism  predicts that the value of the slope parameter b increases as a function of the centre of mass energy like ${\rm b(W_{\gamma p}^2) = b_0 +~ 2 \alpha_{\pom}^\prime ln (W_{\gamma p}^2) }$ where both ${\rm \alpha_{\pom}^\prime}$ and ${\rm b_0 }$, the slope parameter at  
\Wgpm~=~1~\gevtwom, need to be determined experimentally. This mechanism is called `shrinkage' since as  b grows, the area underneath the curve ${\rm e^{b\cdot t}}$ decreases considering a fixed intercept.

As the experimental results become more precise,  it is possible to look for deviations from the pure exponential behaviour ${\rm d\sigma/dt \sim e^{bt}}$ of the t distribution.  An exponential with a quadratic term~\cite{ref:curv} seems to give a good representation of the t dependence for the elastic cross section of many hadronic reactions:
\be 
{\rm 
\frac{d\sigma}{dt} = A\cdot e^{b\cdot t+c\cdot t^2},
}\ee
\noi where c is called  `curvature'. The local slope parameter, defined as:
\be
{\rm
b(t) = \frac{d}{dt}(ln\frac{d\sigma}{dt}) = b + 2ct 
}
\ee is a decreasing  function of t. An example  is given in Fig.~\ref{fig:rho94} for the case of the $\rho^0$ meson. Fits to hadronic scattering data yield to very similar results: b = 11.7 \gevmtwom $\;$ and c = 3.16 ${\rm GeV^{-4}}$ for pp scattering and b = 9.9 \gevmtwom $\;$ and c = 3.47  ${\rm GeV^{-4}}$ for ${\rm \pi p}$ scattering measured  at s = 400 \gevtwom $\;$ in the interval ${\rm 0.02 < |t| <0.66}$ \gevtwom. The common behaviour of the  cross section as  a function of t  is due to the very similar  hadronic form factors for pion, proton and photon.

\begin{table}
\begin{center} \begin{tabular}{|c|c|c|c|c|} \hline
 Reaction             & Collabor. & $\sigma(\mu$b)  &
b(GeV$^{-2}$)&${\rm r^{04}_{00}}$ \\ \hline $\gamma p \ra \rho^0 p$ &
ZEUS &14.7 $\pm$ 0.4 $\pm$ 2.4 & 10.4 $\pm$ 0.6 $\pm$ 1.1 & 0.055
$\pm$ 0.028 \\ 
$\gamma p \ra \rho^0 p$ & ZEUS$_{\rm LPS}$ &5.8 $\pm$ 0.3
$\pm$ 0.7 & 9.9 $\pm$ 0.8
$\pm$ 1.1 & \\
 $\gamma p \ra \rho^0 p$ & H1 & 9.1 $\pm$ 0.9 $\pm$ 2.5 &
10.9 $\pm$ 2.4 $\pm$ 1.1 & -0.11 $\pm$ 0.12 \\ 
$\gamma p \ra \rho^0 X$ & ZEUS$_{\rm LPS}$ &  & 5.3 $\pm$ 0.8
$\pm$ 1.1 & \\
$\gamma p \ra \omega p$
& ZEUS &1.2 $\pm$ 0.1 $\pm$ 0.2 & 10.0 $\pm$ 1.2 $\pm$ 1.4 & 0.11 $\pm$
0.08 \\ $\gamma p \ra \phi p$ & ZEUS &0.96 $\pm$ 0.19 $\pm$ 0.2 & 7.3
$\pm$ 1.0 $\pm$ 0.8 & -0.01 $\pm$ 0.04  \\ \hline
\end{tabular}\end{center} 
\caption{\protect\footnotesize Summary of experimental results on the measurements of ${\rm \sigma_{tot}, \; b \; and\; r_{00}^{04}}$ in photoproduction. The b values are determined using a single exponetial fit  ${\rm d\sigma/dt \sim e^{bt}}$. The cross section ${\rm \gamma p \ra \rho^0 p}$ measured by the LPS is for a restricted t range.} 
\label{tab:bphp}
\end{table} 


\bfm
\begin{center}
\leavevmode
\hbox{%
\epsfxsize = 4in
\epsffile{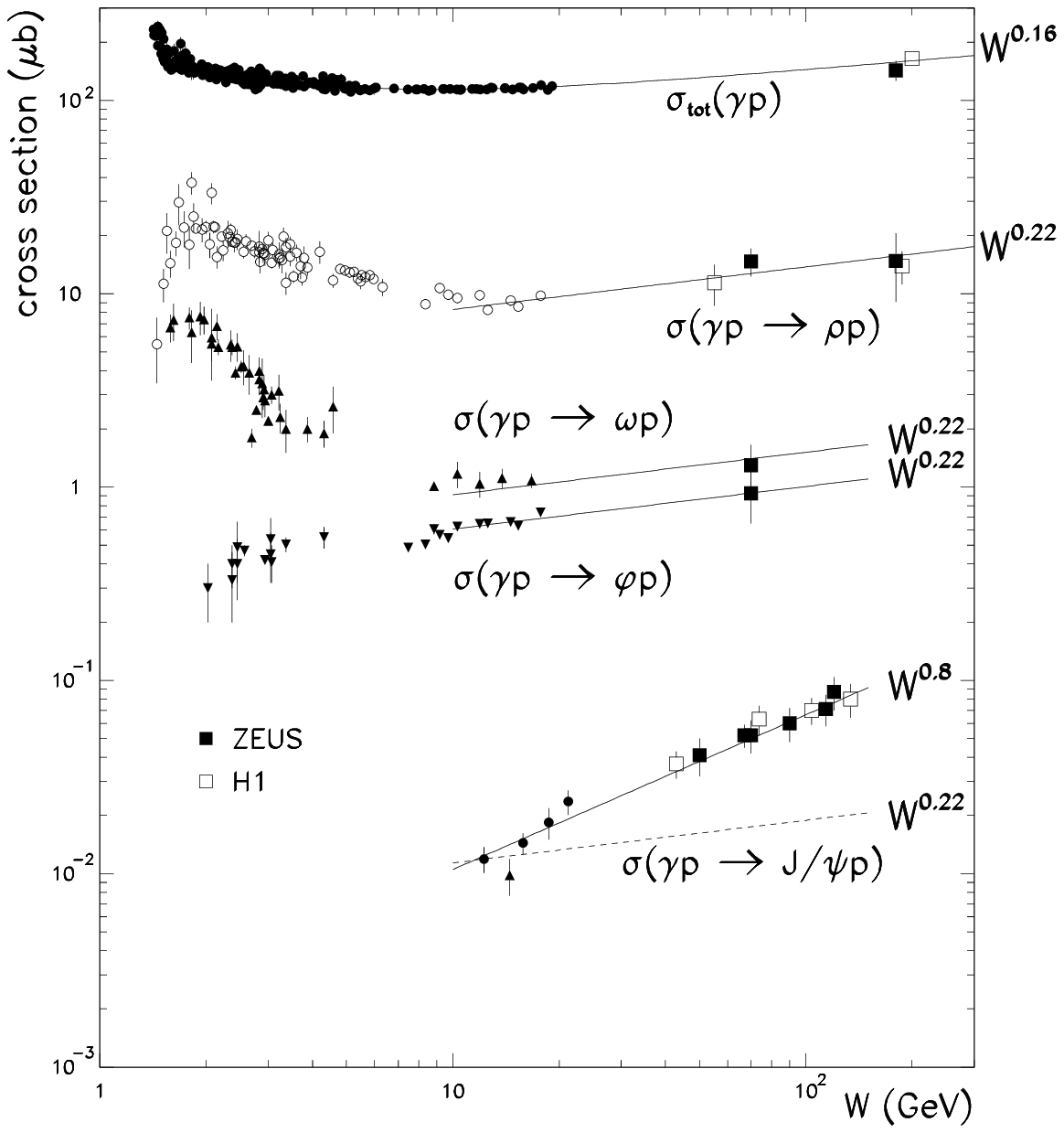}}
\end{center}
\lm{{\protect\footnotesize Total and elastic vector meson photoproduction measurements as a function of \Wgpm. The curve overlapped to ${\rm \sigma_{tot}}$ is the DL parametrization $\Wgpm^{0.16}$. The other lines are curves of the form \Wgpm$^{0.22}$ and  \Wgpm$^{0.80}$. }}
\label{fig:vm_photo_cs}
\end{figure}

The parameter b has been recently measured by the ZEUS collaboration
 in $\rho^0$
photoproduction using data
from the  Leading Proton Spectrometer~\cite{ref:lps} . This is the
first diffractive cross section measurement at HERA in which the
forward scattered  proton is detected and its momentum measured. This
makes possible 
a direct determination of the squared four-momentum t exchanged at
the  proton vertex. The LPS consists of silicon $\mu$-strip detectors placed close to
the proton beam  by means of rentrant Roman pots and
detects forward going protons scattered at angles $\le$1 mrad. The
momentum of the proton is measured using the elements
(quadrupoles and dipoles) of the proton beam line, and it is
reconstructed with a resolution of ${\rm \Delta p/p \simeq }$0.3\% at 
$p\simeq 820$ GeV/c. 
The total systematic error on the
measurement of b in this analysis is 11\%, the main source being 
the uncertainty on the acceptance ($\sim 7$\%), and the uncertainty
coming from the unfolding of the beam transverse momentum spread
($\sim 7$\%). It should be noted that the uncertainty coming from the
proton dissociation background is negligible,  when compared to
analyses which do not make use of the LPS: for LPS tagged events the
contamination has been estimated to be ${\rm 0.21\pm 0.15}$\% while a previous ZEUS result estimated the contamination to be ${\rm 11\pm 6}$\%. 

\noi Tagging with the LPS a leading proton with a value of \xl$<$0.97 has also allowed
to select a clean sample of photoproduction double diffractive $\rho^0$ events, ${\rm \gamma p \ra \rho^0 X}$. Using the transverse momentum from the decay pions, the slope parameter b has been determined to be
${\rm  b_{LPS}^{\gamma p \ra \rho^0 X} =  5.3 \pm 0.8 \pm 1.1}$ \gevmtwom. 
Fig.~\ref{fig:bslope} and Table 3 show the results for both single and double diffraction.

Fixed target experiments showed, at much lower \Wgpm, that  vector mesons retained the  helicity of the photon (s-channel helicity conservation, SCHC). This effect was also investigated at HERA. The results can be expressed in terms of the ${\rm r^{04}_{00}}$ spin-density matrix element which gives the probability for the meson to have zero helicity. As shown in the Tab.~\ref{tab:bphp}, all the measurements are consistent with a zero value for  ${\rm r^{04}_{00}}$, as required by SCHC.

$$\;$$

\bfm
\begin{center}
\leavevmode
\hbox{%
\epsfxsize = 4in
\epsffile{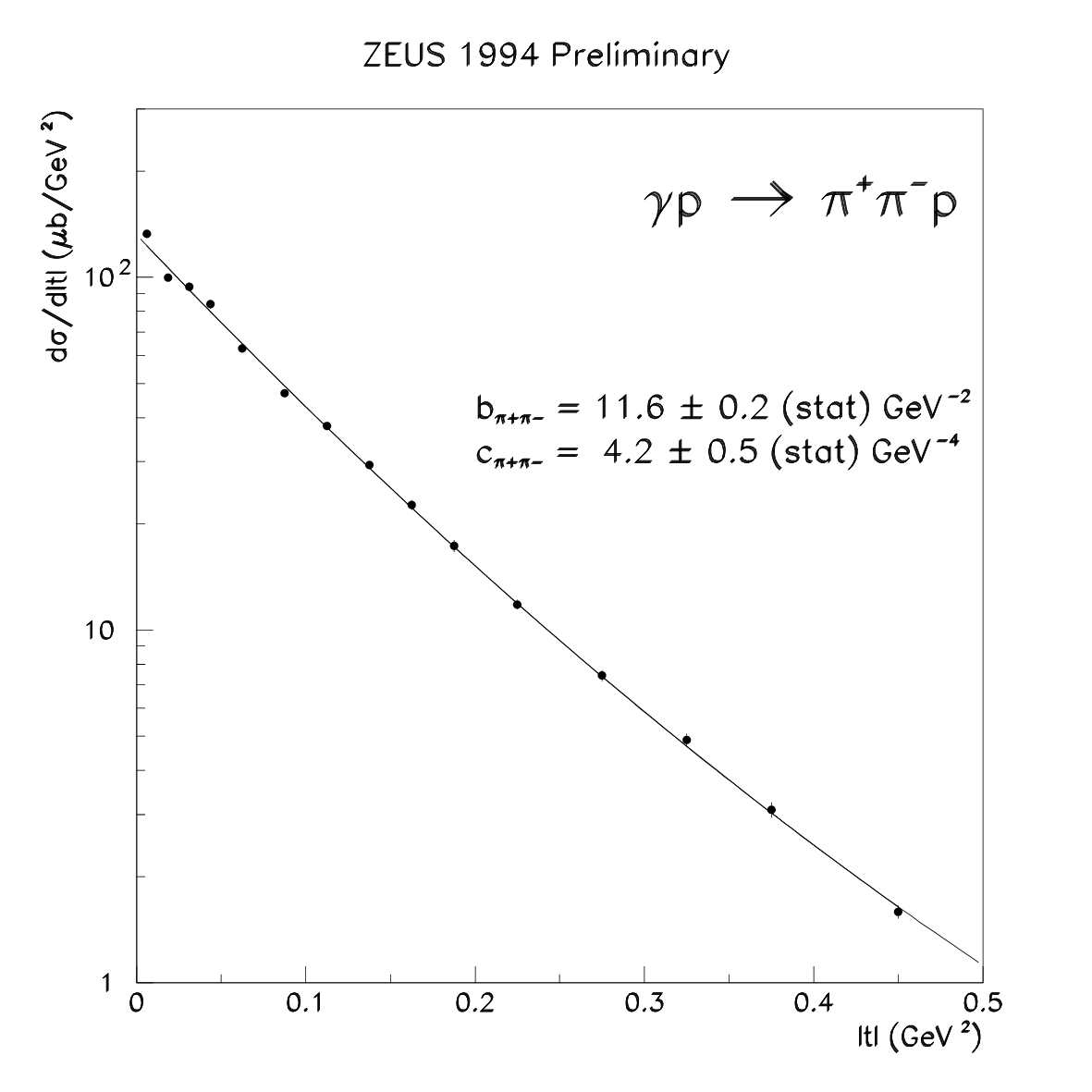}}
\end{center}
\lm{{\protect\footnotesize  Measurement of the slope parameter  for  the reaction ${\rm \gamma p \ra \rho^0 p}$. A quadratic function in t, ${\rm  \frac{d\sigma}{dt} \propto e^{bt+ct^2}}$ was used in the fit.    } }
\label{fig:rho94}
\end{figure}

\bfm
\begin{center}
\leavevmode
\hbox{%
\epsfxsize = 4in
\epsffile{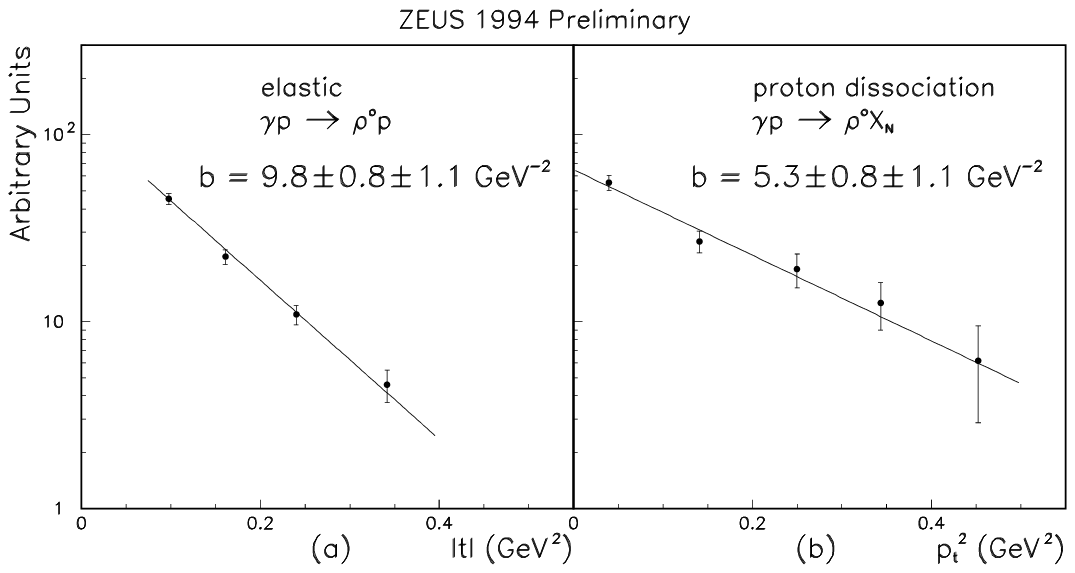}}
\end{center}
\lm{{\protect\footnotesize (a) Slope parameter b for the reaction 
 ${\rm \gamma p \ra \rho^0 p}$ as obtained from the measurement of the scattered proton. The value b=9.8 has been modified to b=9.9 in the final analysis. (b) Slope parameter b for the reaction ${\rm \gamma p \ra \rho^0 X}$ obtained by  tagging a low energy proton in the LPS and measuring ${\rm p_t^2}$ in the tracking chamber.}}
\label{fig:bslope}
\end{figure}

\subsection{Vector meson production with a hard scale}

In contrast to the previous results,  the cross sections for $J/\psi$
photoproduction and  light vector meson production at high \qtwom show
a significant rise
 with \Wgpm. In particular, for the  $J/\psi$ case the rise is clearly visible within  the range of HERA data while for the light vector mesons the rise is observed in comparison with lower energy  data.
Fig.~\ref{fig:vm_photo_cs} and  Fig.~\ref{fig:sigrho}  show the effect for 
the $J/\psi$ and $\rho^0$ case. The  rise   is 
inconsistent with the ${\rm \Wgp^{0.22}}$ dependence used in the parametrizations of low energy hadronic data. The measured behaviour can be described instead by
perturbative  QCD models if a rise at small \xbj of the gluon  momentum density in the
proton is assumed.  Both the shape of the rise and the normalisation
could in principle  be used to discriminate between models of the gluon distributions 
but
since the latter suffers from large theoretical  uncertainty,
 only the shape is used.  Fig.~\ref{fig:rho_vs_x} shows the experimental
 results and the expectation based on different gluon parametrizations for the $\rho^0$
cross section as a function of \xbj in four different \qtwom bins.  
The comparison is still 
dominated by experimental errors but since the parametrizations are 
quite different,  potentially this approach can be an invaluable tool
to rule out many of the current options.
\bfm
\begin{center}
\leavevmode
\hbox{%
\epsfxsize = 4in
\epsffile{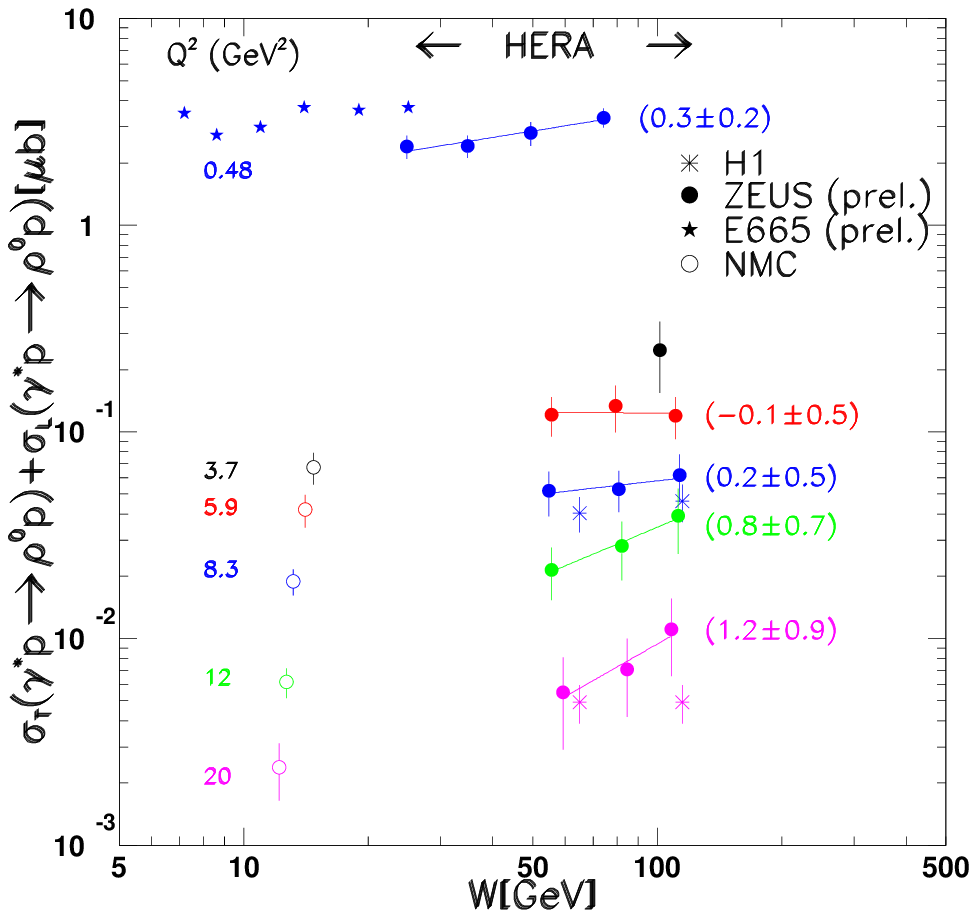}}
\end{center}
\lm{{\protect\footnotesize ${\rm \gamma p \ra \rho^0 p }$ cross section as a function of \Wgpm for different ${\rm <Q^2>}$ values (as indicated in the picture). The lines are the results of fits to the form ${\rm \sigma \propto \Wgpm^k}$, the values of k are shown in parenthesis.}}
\label{fig:sigrho}
\end{figure}
$$ \;$$
\bfm
\begin{center}
\leavevmode
\hbox{%
\epsfxsize = 4in
\epsffile{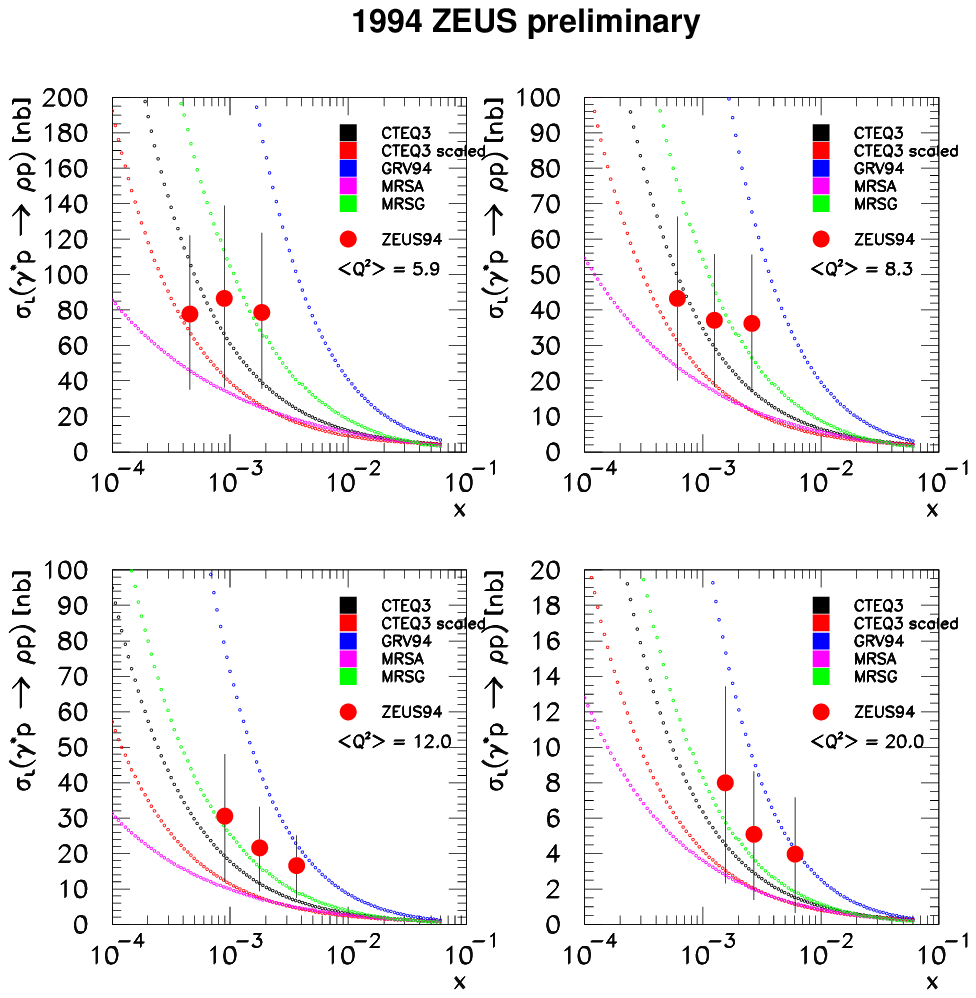}}
\end{center}
\lm{{\protect\footnotesize  ${\rm \gamma p \ra \rho^0 p }$ cross section as a function of \xbj for different values of ${\rm <Q^2>}$. The lines are prediction from a calculation based on pQCD using different gluon distributions. }}
\label{fig:rho_vs_x}
\end{figure}

\vspace{-1.5cm}
These results  show that  the cross section for   vector meson production in the presence of a hard scale has a steeper energy dependence than the total hadronic ${\rm \gamma p}$ cross section. 

\subsubsection{Slope parameter b  and ${\rm R=\sigma_L/\sigma_T}$ vs \Wgpm and \qtwom}

The slope parameter b is related to the effective radius of the interaction R by:
\be{\rm
R=\sqrt{R^{2}_{p}+R^{2}_{VM}}\simeq 0.3 \; b^{1/2} \;\; fm 
}
\ee
\noi with ${\rm R_p}$ and ${\rm R_{VM}}$ the proton and  vector meson radius.
\noi Given a value of   ${\rm R_{p}\simeq\; 0.7}$ fm,  the effective vector meson radius in photoproduction, according to  Tab.~\ref{tab:bphp}, changes from ${\rm R_{VM}\simeq 1.1}$ fm for the $\rho^0$ meson to  ${\rm R_{VM}\simeq 0.4}$ fm  for the $\phi$ meson.

The value of  b  varies with  the meson mass, the photon virtuality \qtwom and the square of the 4-momentum transfer t. Fig.~\ref{fig:bvsq2} shows a compilation of the measurements done  by both H1 and ZEUS as a function of ${\rm Q^2+M_{VM}^2}$. The data show a clear trend toward small b values as  ${\rm Q^2+M_{VM}^2}$ increases. Note that some authors~\cite{ref:frank} use an `effective \qtwom' instead of the measured \qtwom to set the scale. The drop of b from b $\simeq$ 10
 to b $\simeq$ 4-5 ${\rm GeV^{-2}}$ implies that the size of the system (the ${\rm \gamma^*\ra \rho^0} $ Pomeron vertex) decreases with  ${\rm Q^2+M_{VM}^2}$ and that for large  ${\rm Q^2+M_{VM}^2}$ we do have a short distance interaction to justify the use of pQCD. The value   b $\simeq$ 4-5 ${\rm GeV^{-2}}$ is approximately
 equal to what is expected from the size of the proton.
\bfm
\begin{center}
\leavevmode
\hbox{%
\epsfxsize = 3in
\epsffile{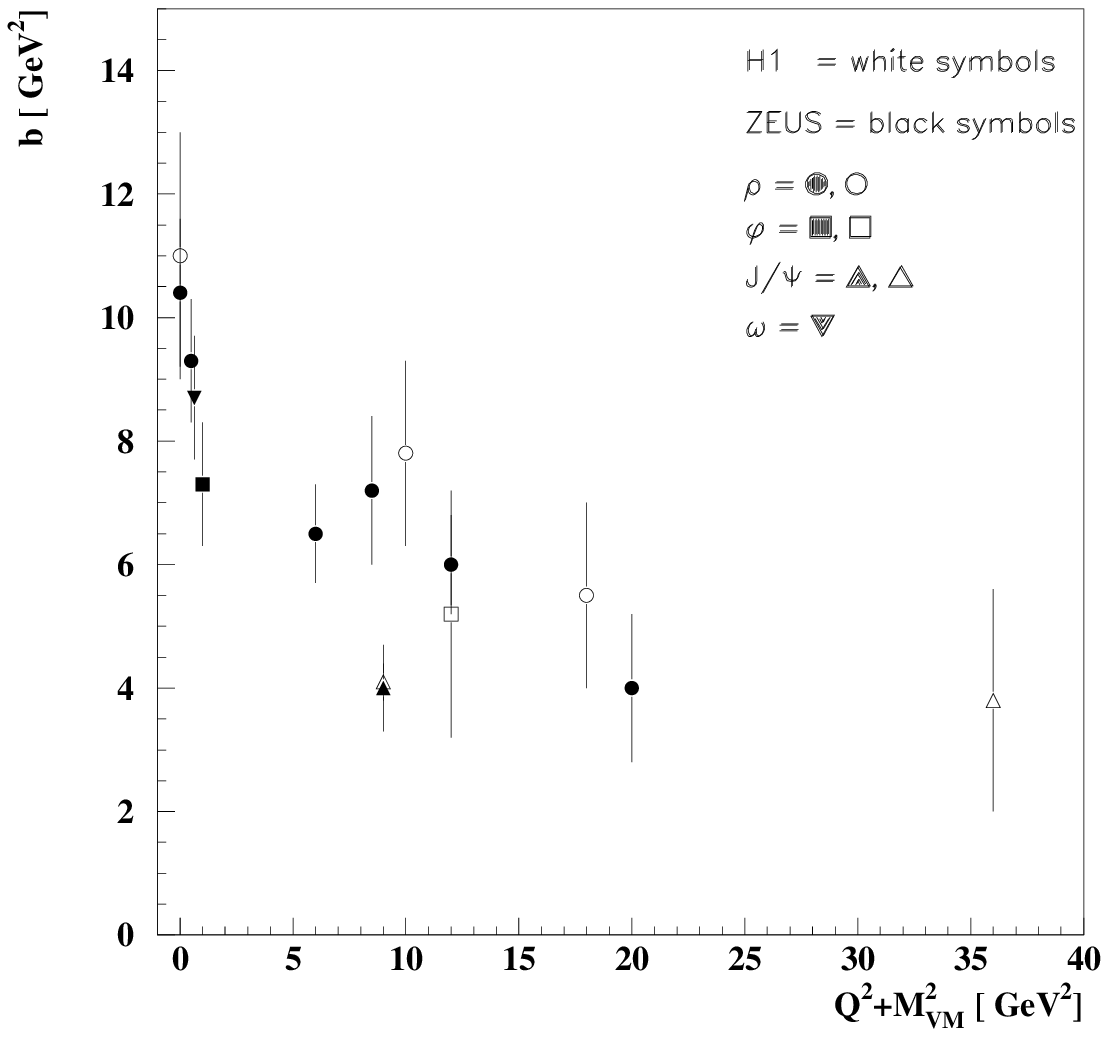}}
\end{center}
\lm{{\protect\footnotesize Exponential slope b for vector meson production as a function of  ${\rm Q^2+M_{VM}^2}$.  }}
\label{fig:bvsq2}
\end{figure}

Both the H1 and ZEUS collaborations have studied  the ratio between the longitudinal and transverse cross section for $\rho^0$ production, ${\rm R = \frac{\sigma_L}{\sigma_T}}$,
  as a function of \qtwom. A compilation of the results is shown in Fig.~\ref{fig:r}.  The photon polarisation, completely transverse at \qtwom=0 \gevtwom, becomes more longitudinal as \qtwom increases. Different QCD calculations have been done (for a review\cite{ref:ry2}). In particular, the  convolution of the ${\rm \gamma^* \ra q\bar{q}}$ diffractive production with the $\rho^0$ meson wave function gives:
\be
{\rm
 R = \frac{\sigma_L}{\sigma_T} \propto \frac{Q^2}{m^2_{\rho^0}},
}\ee
\noi which has a  much too steep dependence with \qtwom. A new approach\cite{ref:ry2} , based on the parton-hadron duality, couples the rise with \qtwom to the gluon distribution anomalous dimension $\gamma$:
\be
{\rm  R = \frac{\sigma_L}{\sigma_\perp} \simeq \frac{Q^2}{M^2_X}(\frac{\gamma}{\gamma+1})^2.
}\ee
\noi  Since $\gamma$ decreases with \qtwom, a less steep dependence is obtained
that seems to fit the data quite well. 
\bfm
\begin{center}
\leavevmode
\hbox{%
\epsfxsize = 3in
\epsffile{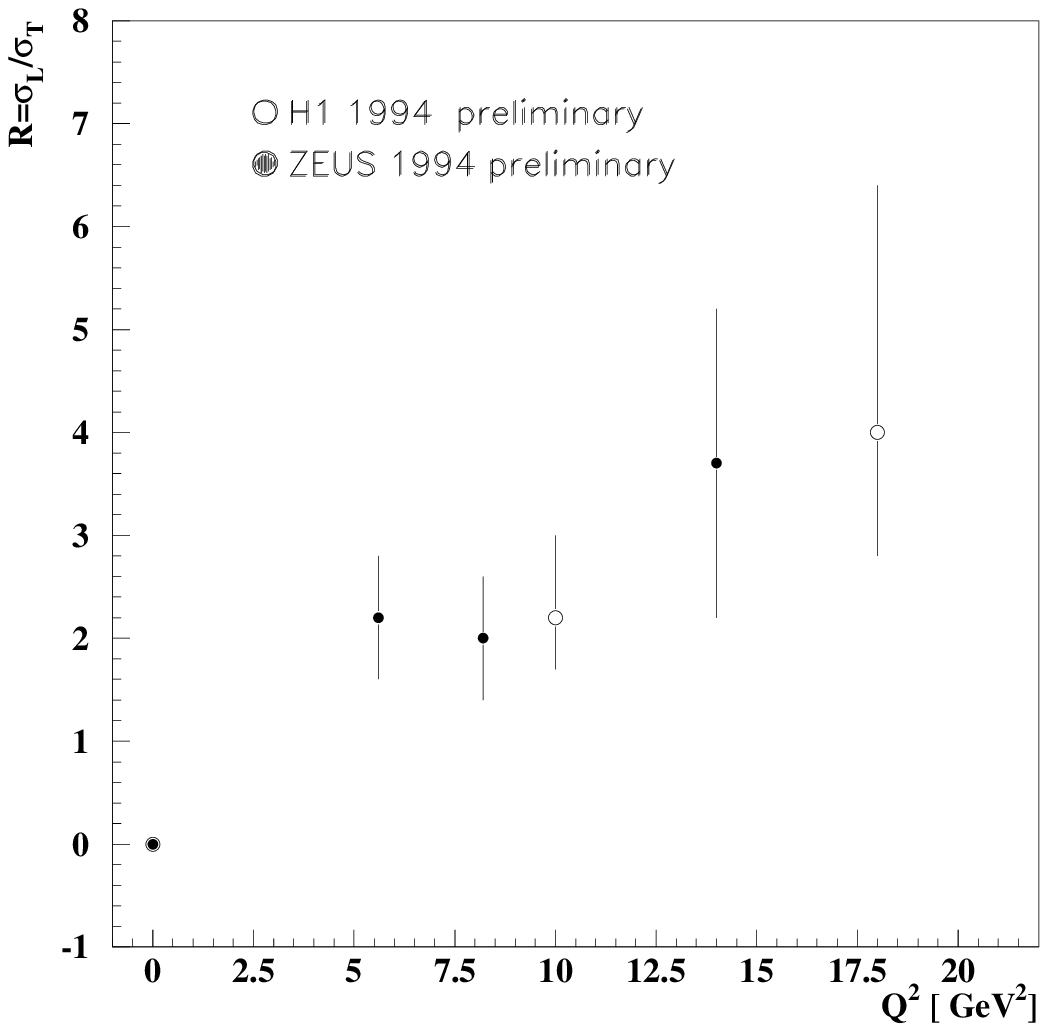}}
\end{center}
\lm{{\protect\footnotesize  Ratio ${\rm R = \sigma_L/\sigma_\perp}$  for the reaction ${\rm \gamma p \ra \rho^0 p}$ as a function of the photon virtuality \qtwom.}}
\label{fig:r}
\end{figure}
Within the current experimental accuracy, R does not seem to depend on \Wgpm.

\subsubsection{Determination of ${\rm \alphapom^\prime}$ at large \qtwom }

\noi As we have seen,  where Regge theory holds, the value of b should increase with energy according to the `shrinkage' mechanism. The HERA data  on photoproduction of  $\rho^0, \phi, \omega$ are consistent with this prediction. At high \qtwom there are no pre-HERA measurements of the `shrinkage' mechanism. According to~\cite{ref:nzsh} , ${\rm \alphapom^\prime \simeq}$ = 0.2 \gevmtwom $\;$. Therefore the value of the slope parameter b should increase by $\sim $ 1.5 \gevmtwom$\;$ going from low energy, \Wgpm $\sim$ 10 GeV, to the HERA regime, \Wgpm $\sim$ 100 GeV, for all exclusive reactions of the kind ${\rm \gamma^* p \rightarrow Vp}$. On the other hand, according to~\cite{ref:hal2} , the value of ${\rm \alphapom^\prime}$ is expected to be  ${\rm \alphapom^\prime \sim 1/Q^2}\;$ for reactions where the hard QCD regime dominates, leading to a flat behaviour of b as a function of W.  

\noi Fig.~\ref{fig:slopew_rho} shows the b values for the $\rho^0$ measurements as a function of \Wgpm at high \qtwom.  The experimental data are still dominated by statistical errors and therefore no meaningful conclusion can be drawn. On the plot, the expected trend of b vs \Wgpm is plotted if a value of ${\rm  \alpha_{\pom}^\prime}$ = 0 or 0.25 is assumed. 

\bfm
\begin{center}
\leavevmode
\hbox{%
\epsfxsize = 3in
\epsffile{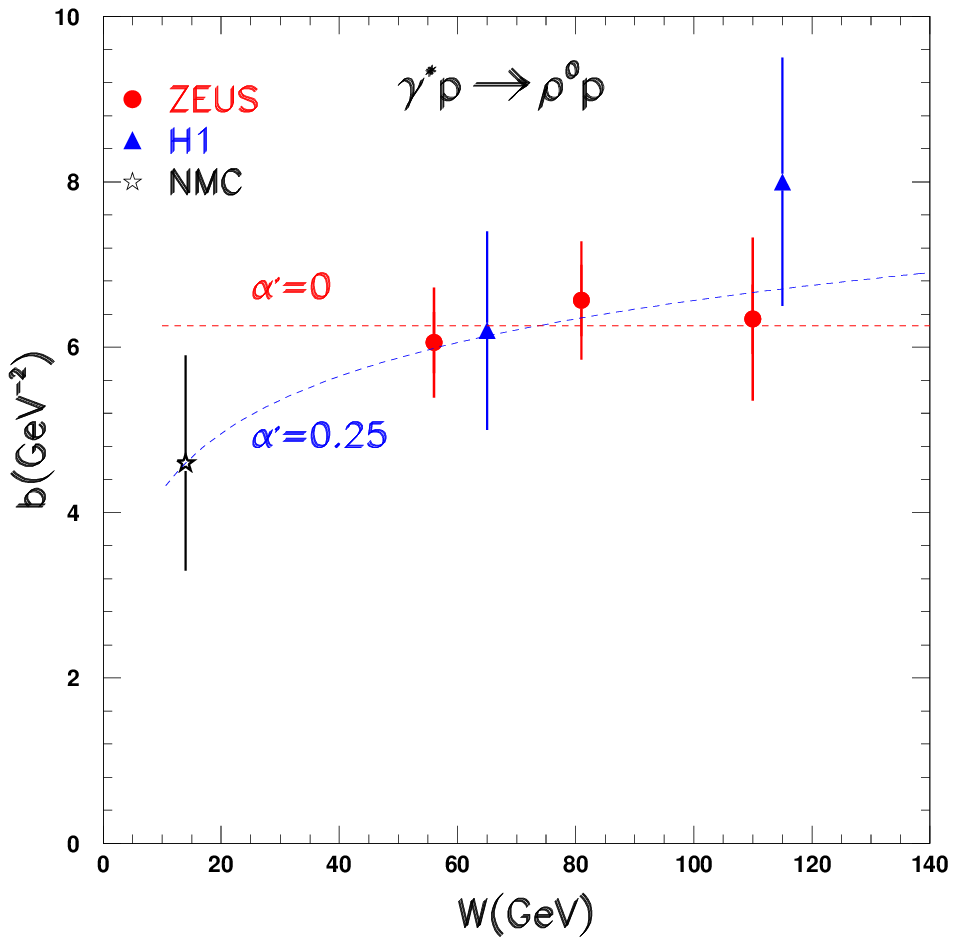}}
\end{center}
\lm{{\protect\footnotesize  Exponential slope b for the reaction ${\rm \gamma p \ra \rho^0 p}$ as a function of \Wgpm.   On the plot, the expected trend of b vs \Wgpm is plotted if a value of ${\rm  \alpha_{\pom}^\prime}$ = 0 or 0.25 is assumed. }}
\label{fig:slopew_rho}
\end{figure}

\subsubsection{Restoration of SU(4) symmetry at high \qtwom} 

 According to the SU(4) flavour symmetry, when ${\rm M^2_V} \;\ll$ \qtwom,  the ratio among cross sections for diffractive vector meson production should depend only on the mesons wavefunction and quark charges~\cite{ref:frank}:
\be
{\rm \rho:\omega:\phi:J/\Psi = [\frac{1}{\sqrt{2}}(u\bar{u}+d\bar{d})]^2:[\frac{1}{\sqrt{2}}(u\bar{u}-d\bar{d})]^2:[s\bar{s}]^2:[c\bar{c}]^2  =  9:1:2:8}.
\ee
\noi Besides, QCD dynamics predicts a slow increase of the relative yield of heavy flavour production at small \xbj which modifies the pure SU(4) prediction:
\be
{\rm \rho:\omega:\phi:J/\Psi = 9:(1*0.8):(2*1.2):(8*3.5)}.
\ee
The HERA results are shown in ~Fig.~\ref{fig:ratio}: at ${\rm Q^2 \sim 0.}$ \gevtwom , SU(4) symmetry is badly broken, with a suppression factor $\simeq$ 4 for ${\rm \phi}$-mesons and $\simeq$ 25 for ${\rm J/\psi}$-mesons while at large \qtwom there is a dramatic increase of both the ${\rm \phi \;\; and \;\;J/\Psi}$ cross section compared to $\rho^0$ meson production. This increase is therefore another indication that the SU(4) symmetry, and therefore perturbative QCD, can be used in these processes at large \qtwom.

\noi Cross section ratios between excited and ground states for a meson are also very important quantities because they depend on the internal dynamics of the \qqbar wavefunction and can help to determine it. First preliminary results from the H1 collaboration\cite{ref:jp2s1,ref:rhop1} , in  agreement with the expectation of~\cite{ref:frank} , are: \\
$${\rm
 \frac{\sigma^{\rho^\prime}}{\sigma^{\rho}} =    0.36\;\pm\;0.07\;\pm0.11 {\rm \; at \;\qtwom\; =\; 4-50\; GeV^2  }}$$ 
$${\rm \frac{\sigma^{\Psi(2S)}}{\sigma^{\Psi}} =    0.16\;\pm\;0.06 {\rm\; at\; Q^2 \;=\; 0\; GeV^2 }.}$$ 

\bfm
\begin{center}
\leavevmode
\hbox{%
\epsfxsize = 4in
\epsffile{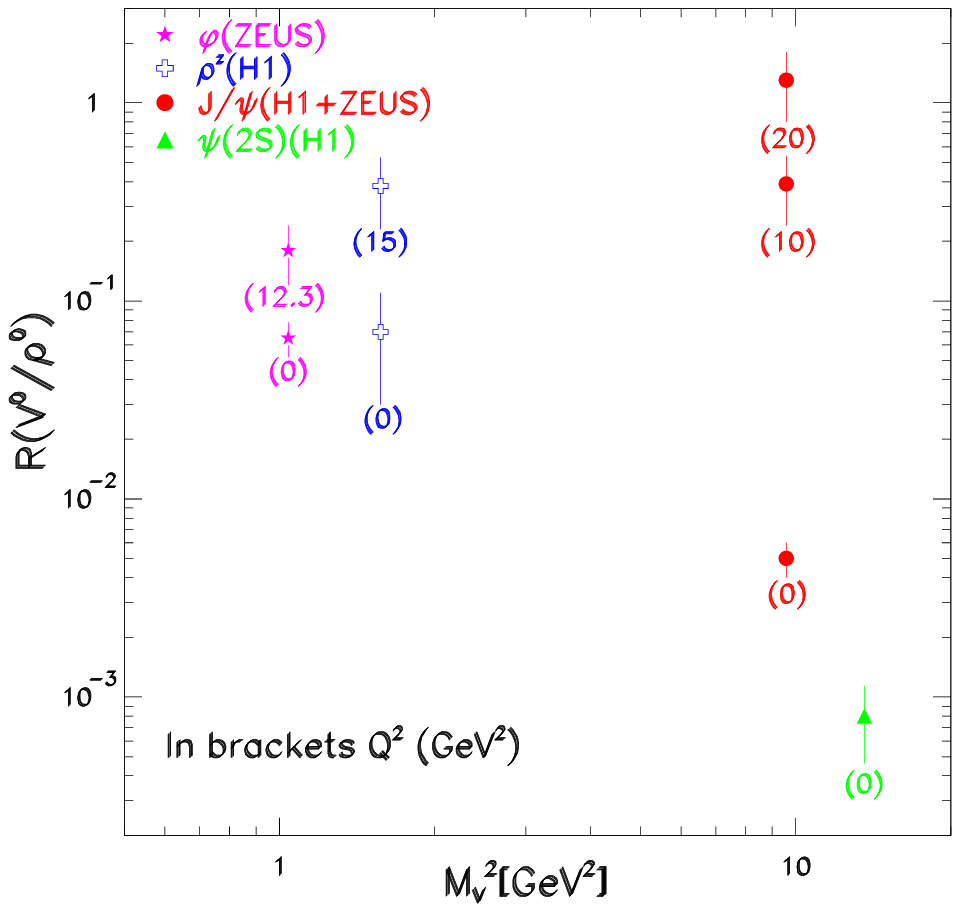}}
\end{center}
\lm{{\protect\footnotesize Ratio ${\rm R = \sigma^{\gamma p \ra V p}/\sigma^{\gamma p \ra \rho^0 p}}$ as a function of the  vector mesons mass squared ${\rm M^2_V}$ at different values of the photon virtuality \qtwom (indicated by the number in parenthesis).}}
\label{fig:ratio}
\end{figure}


\section{Photon diffraction: \rm{$ \gamma p$} \ram X Y}
Single and double photon diffraction include all pomeron mediated reactions of the kind:
\bc
${\rm \gamma  p \longrightarrow X Y}$
\ec
\noi where X  is not a vector meson and Y is either a proton or an excited state. These reactions can be divided into two large groups depending on whether a hard scale is present in the scattering process.

\subsection{Experimental signature and selection methods for ${\rm \gamma  p \longrightarrow X Y}$}
As it was shown in Section (\ref{sec:expsig}),  diffractive events generally have a rapidity gap and a leading barion in the final state.  Several  selection methods have been used by both the H1 and ZEUS collaborations exploiting their own detectors. In the following the  four  most significant  methods are presented.

\subsubsection{Maximum pseudorapidity ${\rm \eta_{max}}$ (ZEUS,H1)}
 
At HERA, following the first papers on the subject~\cite{ref:zrap1,ref:h1eta} , a cut on   the  pseudorapidity of the most forward\footnote{As it was said before, in the HERA convention the proton travels along the z-axis in the positive direction} energy deposit in an event  has been used to separate diffractive from non diffractive events. This cut selects as diffractive events all those events whose most forward energy deposit has a  rapidity  less than 1.5, equivalent to require a visible rapidity gap of at least 2.9 unit in the forward direction (ZEUS case). 
This cut, however, puts strong limitations on the type of events that are selected since it reduces the pseudorapidity interval available for the fragmentation of the system \mx to  ${\rm  \Delta\eta \simeq}$ 4.5 - 5.5. Since a system with mass \mx covers a pseudorapidity interval ${\rm  \Delta\eta \simeq ln(\frac{M_X^2}{m_p^2})}$ with ${\rm m_p}$ the proton mass,  only  masses up to \mx$\sim$ 10-15 GeV are therefore selected.


\subsubsection{Largest rapidity gap (H1)}

For each event, the largest rapidity gap is identified, Fig.~\ref{fig:difXYdef}. This gap defines two systems,  X and Y with masses $\mx$ and ${\rm M_Y}$. If:
\bi
\item [a)] ${\rm   \xpom = \frac{M_X^2+Q^2}{W_{\gamma p}^2+Q^2}\; < \; 0.05 }$
\item [b)] ${\rm  M_Y\; <\; 1.6 \; GeV},$
\ei
the event is accepted in the diffractive sample. This selection is based on the H1 detector ability of measuring hadronic activity up to ${\rm \eta \sim 3.4 \;\; \Rightarrow \xpom < 0.05}$ and vetoing activity in the region ${\rm 3.4 < \eta < 7.5 \;\; \Rightarrow M_Y < 1.6}$ GeV. 

\noi The requirement a) ensures that only a small fraction of the initial proton longitudinal momentum is present in the detector while b) forces the existence of a rapidity gap in the final state. It is important to notice that this selection criterium does not make any assumption on the nature of the interaction but defines a cross section for all events that are selected by a) and b).

\bfm
\begin{center}
\leavevmode
\hbox{%
\epsfxsize = 2.5in
\epsffile{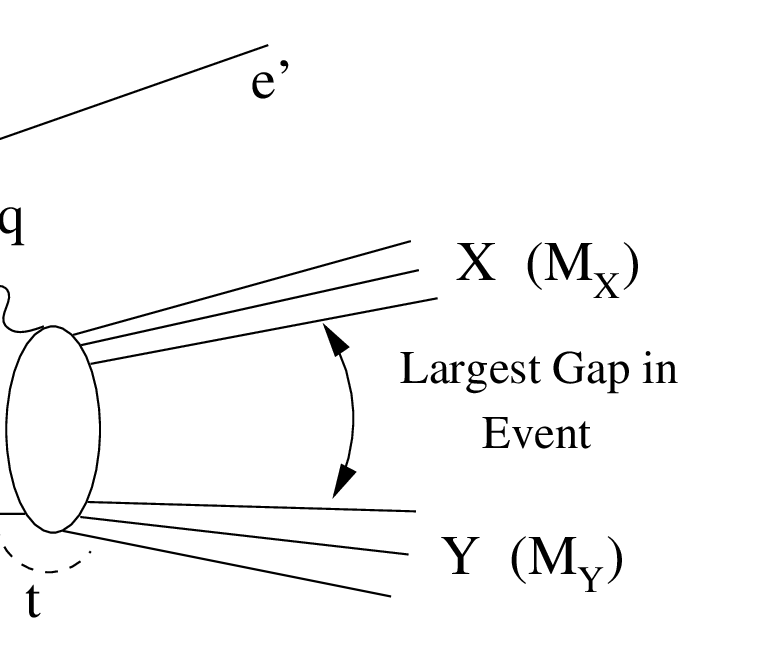}}
\end{center}
\lm{{\protect\footnotesize  Schematic illustration of the selection method used to define a diffractive cross section  by the H1 collaboration.}}
\label{fig:difXYdef}
\end{figure}

\subsubsection{Leading proton measurement (ZEUS\_LPS)}

The cleanest way to identify diffractive events is to tag a scattered proton with a very high fraction \xl of the initial proton momentum.  From the leading proton momentum, both t and the hadronic mass \mx can be computed:
\be{\rm
t\simeq -\frac{p_t^2}{x_L}
}\ee
\be
{\rm
\mxtwo = \Wgp^2\cdot (1-x_L).
}
\ee

\noi Fig.~\ref{fig:xl_lps} shows the fraction of DIS events with a leading proton as a function of \xl for ${\rm 5 < Q^2 < 20 }$ \gevtwom, ${\rm 45 < \Wgpm <225}$ GeV and ${\rm (1-x_L)^2/x_L < |t| < 0.5}$ \gevtwom. Comparing this figure with Fig.~\ref{fig:xl}, two different components  can be easily identified: the diffractive peak due to \pom $\;$ exchange at \xl$\sim$ 1 and the continuum due to double dissociation, reggeon exchange and non-diffractive DIS scattering rising below \xl $\sim$ 0.9. 
\bfm
\begin{center}
\leavevmode
\hbox{%
\epsfxsize = 5in
\epsffile{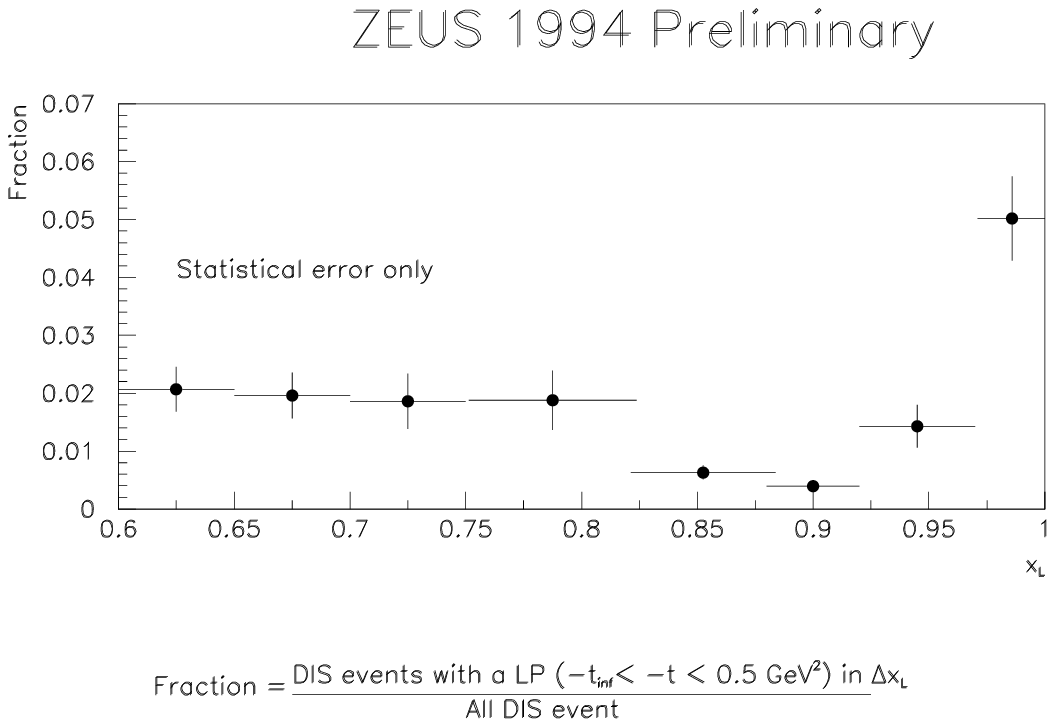}}
\end{center}
\lm{{\protect\footnotesize Fraction of DIS events with a leading proton as a function of \xl.
}}
\label{fig:xl_lps}
\end{figure}
Just below \xl $\sim$ 1, 
the distinction between diffractive and non diffractive events becomes unclear. From a fit to the \xl spectrum in pp scattering~\cite{ref:gou} , the value \xl$\sim$ 0.9 has been used to identify the point where the pomeron and non pomeron contributions are roughly of the same magnitude. The ZEUS collaboration, in order to select a very pure diffractive sample, decided to use only protons with \xl$>$ 0.97, well within the diffractive peak. Unfortunately, due to the limited LPS acceptance, the number of events with a tagged leading proton is small.
Note that the LPS acceptance, at \xl$>$0.95, starts at ${\rm t \sim 0.07 GeV^2}$. 

\subsubsection{Ln(\mx) distribution (ZEUS\_MX)}
 This method~\cite{ref:zeusmx} of separating the diffractive and non diffractive contributions is based on their very different ${\rm M_X^2}$ distributions.

\noi Non-diffractive events, assuming uncorrelated particle emission, have  an exponential  fall-off of the ln ${\rm M^2_X}$ distribution:
\be
{\rm 
\frac{d{\cal N}^{nondiff}}{d\ln M^2_X} = 
c\, \exp (b \ln M^2_X)
}\ee
while diffractive events have a constant value in  the ${\rm \ln M^2_X}$
 distribution:
\be
{\rm
 \frac{dN}{d\mxtwo} \sim \frac{1}{\mxtwo} \Longrightarrow \frac{dN}{d\ln \mxtwo}\; =\; const.
}
\ee
The diffractive sample is therefore defined as the excess contribution in the ln ${\rm M_X^2}$ distribution above the exponential fall-off of the non-diffractive peak. In bins of \Wgpm, \qtwom and \mx,  a fit in the form:
\be
{\rm
\frac{dN}{d\ln \mxtwo} = D +  c\, \exp (b\ln {\mxtwo})
}\ee
is performed allowing the evaluation of the diffractive component. An example of the ln ${\rm M_X^2}$ distribution is given in~Fig.~\ref{fig:wolf1}.

\bfm
\begin{center}
\leavevmode
\hbox{%
\epsfxsize = 3in
\epsffile{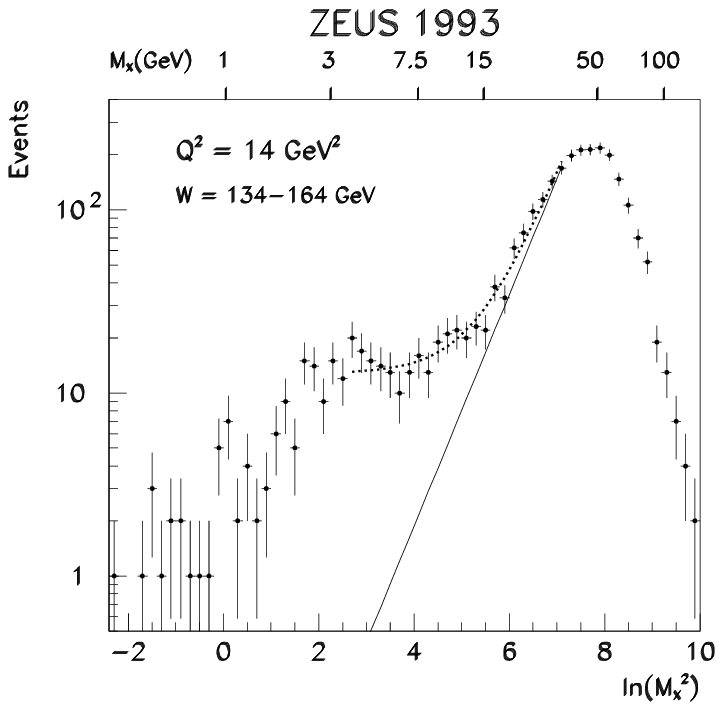}
}
\end{center}
\lm{{\protect\footnotesize ln ${\rm M_X^2}$ distributions for ${\rm \gamma^* p}$ scattering. Diffractive events are identified as the excess contribution above the exponential fall-off of the non-diffractive peak.}}
\label{fig:wolf1}
\end{figure}

\subsection{Determination of $\alphapom$ and test of factorization}


The assumption of factorization implies that the pomeron 
structure is independent of the process of emission and that the pomeron flux is the same in all diffractive processes. Its dependence on $\alphapom$ is given by:
\be
{\rm 
 f_{\pom/ p}(\xpom) \propto (\frac{1}{\xpom})^a,
}
\ee

\noi with ${\rm a =  2\alphapom(0)-1}$. If factorization holds, the same value of  ${\rm \alphapom(0)\simeq 1.08 }$  measured  in many hadronic reactions should also control the pomeron flux  at HERA.

\bfm
\begin{center}
\leavevmode
\hbox{%
\epsfxsize = 3in
\epsffile{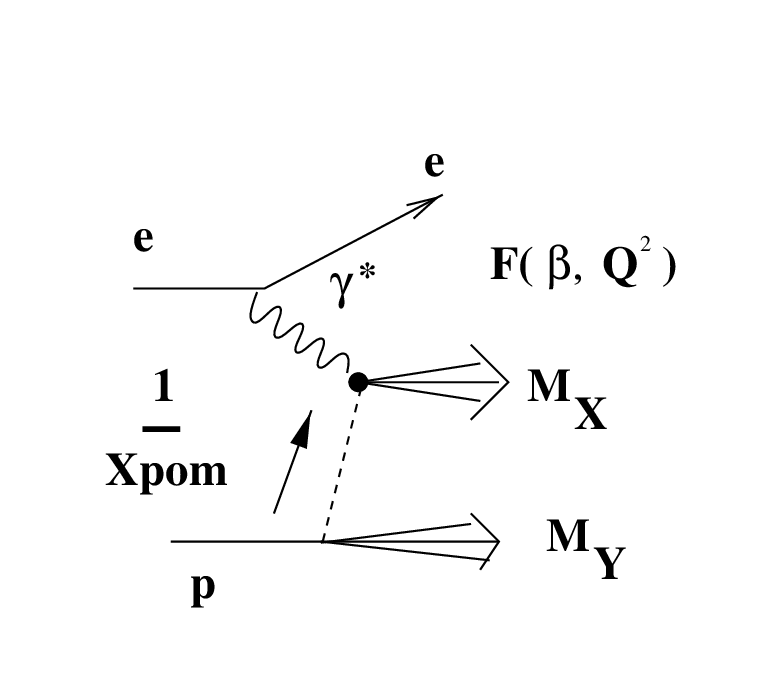}}
\end{center}
\lm{{\protect\footnotesize Diffractive ep scattering according to factorization: the flux of pomeron and the pomeron structure function are universal quantity that can be determined separately.}}
\label{fig:fact}
\end{figure}

\subsubsection{Determination of $\alphapom$ at \qtwom $\simeq$ 0 \gevtwom}
The value of ${\rm \alphapom(0)}$, as shown in Section 2,  can be measured directly from  the behaviour of ${\rm \sigma_{tot}^{\gamma p }(W)}$ as a function of the centre of mass  energy \Wgpm. Results from both the H1 and ZEUS collaborations are consistent with a value of ${\rm \alphapom(0) \simeq}$ 1.08  (see Section (\ref{sec:regge})).

A second method to determine ${\rm \alphapom(0)}$ is based on the behaviour of the differential cross section ${\rm \frac{d^2\sigma}{d|t|dM_X^2}}$  as calculated in the Regge formalism for the triple pomeron diagram:
\be
{\rm 
\frac{d^2\sigma}{d|t|d\mxtwo} \propto ( \frac{1}{\mxtwo})^{\alphapom(0)} \cdot e^{(b_o+2\alphapom^{\prime} ln(\frac{W_{\gamma p}^2}{M_X^2}))\cdot|t|}.
}
\ee

\begin{table}
 \bc \begin{tabular}{| c| c| c|} \hline
Collaboration & \mx interval $[$ GeV $]$ &  $\alphapom(0)$ \\ \hline 
 ZEUS & ${\rm 8\; <\; \mx\; <\; 24}$ & ${\rm  1.14 \;\pm\; 0.04\;(stat)\; \pm\; 0.08 \;(syst)}$ \\ \hline 
 H1 &${\rm 3\; <\; \mx\; <\; 24}$ &  ${\rm  1.11 \;\pm\; 0.02\;(stat)\; \pm\; 0.07 \;(syst)}$ \\ \hline
\end{tabular}
\caption{\pf ${\rm \alphapom(0)}$ values in diffractive photoproduction as determined by the H1 and ZEUS collaborations from a fit to ${\rm \frac{d^2\sigma}{d|t|dM_X^2}}$.}
\label{tab:phpalpha}
\ec
\end{table}
\noi In  Tab.\ref{tab:phpalpha}, the results from the H1~\cite{ref:hmx} and ZEUS~\cite{ref:zmx} collaborations are presented. Both  results  suggest a value for ${\rm \alphapom(0)}$ consistent with Regge phenomenology as already indicated by  ${\rm \sigma_{tot}^{\gamma p }(W)}$ and support the hypothesis that the same `soft Pomeron' used to describe the high energy behaviour of hadron-hadron scattering is also responsible for diffractive
photoproduction at HERA.

\subsubsection{Determination of $\alphapom$ at large \qtwom }

\noi At large \qtwom, by analogy with standard deep inelastic scattering,
the differential cross section for deep inelastic diffractive scattering can be written as:
\be
{\rm
  \frac{d^4\sigma_{diff}}{dQ^2 d\beta d\xpom dt} = \frac{2 \pi
    \alpha^2}{\beta Q^4} \; (1+(1-y)^2) \;
  F_2^{D(4)}(Q^2,\beta,\xpom,t).
\protect\label{eq:f2}
}
\ee
Using now the assumption that factorization is valid, the t and \xpom$\;$ dependence can be separated from the dependence on $\beta$ and \qtwom:
$$ {\rm F_2^{D(4)}(Q^2,\beta,\xpom,t) = f(x_{I\!\!P},t)\cdot F_2^{D(2)}(Q^2,\beta)}.$$
Integrating the pomeron flux  over t and writing the dependence on \xpom $\;$explicitly, ${\rm  F_2^{D(4)}(Q^2,\beta,\xpom,t)}$ becomes:

\be
 {\rm    F_2^{D(4)}(Q^2,\beta,\xpom,t) \;\; \Rightarrow \;\; (\frac{1}{\xpom})^{a} \cdot  F_2^{\pom}(Q^2,\beta)}.
\protect\label{eq:flux}
\ee

\noi Following eq.~\ref{eq:f2},\ref{eq:flux}, the determination of the DIS diffractive cross section in \xpom bins provides a method to measure $\alphapom$ as a function of \qtwom and $\beta$ and to test whether  factorization holds.

\par
\vspace{0.5cm}
{\bf H1 determination of $\alphapom$} \nl
\bfm
\begin{center}
\leavevmode
\hbox{%
\epsfxsize = 2.5in
\epsffile{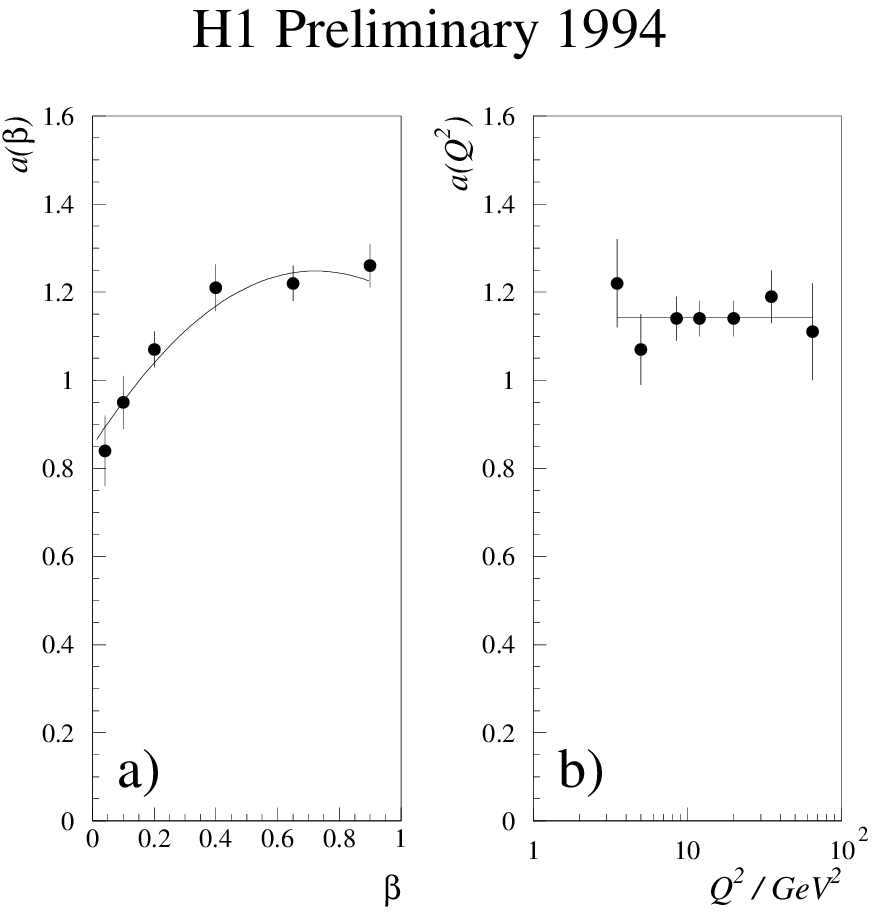}}
\end{center}
\lm{{\protect\footnotesize Results for the value of a when  ${\rm  F_2^{D(2)}(\beta,Q^2)}$ is fitted to the form ${\rm (\frac{1}{\xpom})^{a(\beta)}}$. Statistical and systematic errors are added in quadrature. }}
\label{fig:xpomnbq2}
\vspace{-0.2cm}
\end{figure}
\noi Fig.~\ref{fig:x22} shows the quantity ${\rm \xpom \cdot F_2^{D(2)}(\beta,Q^2)}$ for different \qtwom and $\beta$ bins as a function of \xpom.  The data~\cite{ref:hf2d} were fitted with the function:
$$ {\rm F_2^{D(3)} =  (\frac{1}{\xpom})^{a} \cdot A(\beta, Q^2) } $$
\noi in  each ${\rm \beta \; or \; Q^2}$ interval. The points
 clearly show a change in slope  going from low to high values of $\beta$  while no dependence is seen with \qtwom, Fig.~\ref{fig:xpomnbq2}. This analysis, therefore, shows that there is a change in the value of n as a function of \xpom and ${\rm \beta}$. This experimental factorization breaking, however, 
does not uniquely indicate a change in the pomeron flux  but it might also be explained in terms of a presence in  the data of  a `non-pomeron'  component. If  a fit using a meson and a pomeron component is performed, Fig.\ref{fig:x22},
$$ {\rm F_2^{D(3)} = F_2^{I\!\!P}(\beta,Q^2) \cdot \xpom^{-a}
+ C_M \cdot F_2^{M}(\beta,Q^2) \cdot \xpom^{-n_2}}, $$

\noi then a single value for `a' gives a good description of the data (preliminary results):
$${\rm a\; =\; 1.29\;\pm\;0.03\;(stat)\;\pm\;0.06(syst)\;\pm\;0.03\;(model)  }$$
$${\rm n_2\; =\; 0.3\;\pm\;0.3\;(stat)\;\pm\;0.6\;(syst)\;\pm\;0.2\;(model)}$$
\noi with ${\rm \chi^2/ndf = 170/156}$. The value obtained for ${\rm n_2}$ is consistent with what is expected from meson exchange.  From a, a value for $\alphapom$ averaged over the unmeasured t distribution, $\alphapombar$, can be obtained,
$\alphapombar = \frac{a+1}{2} $:
\vspace{-0.1cm}
$${\rm  \alphapombar^{H1}  = 1.15\pm0.02 \;(stat) \;\pm0.04 \;(syst)}.$$

To check consistency between results obtained with this method and a previous
H1 analysis~\cite{ref:h1eta} ,
the measurement of ${\rm \alphapom}$ using a single component has been performed also over the same kinematical range used in~\cite{ref:h1eta} , obtaining a result
that is compatible within statistical errors with the old one.

$$\;$$
\newpage

\includegraphics{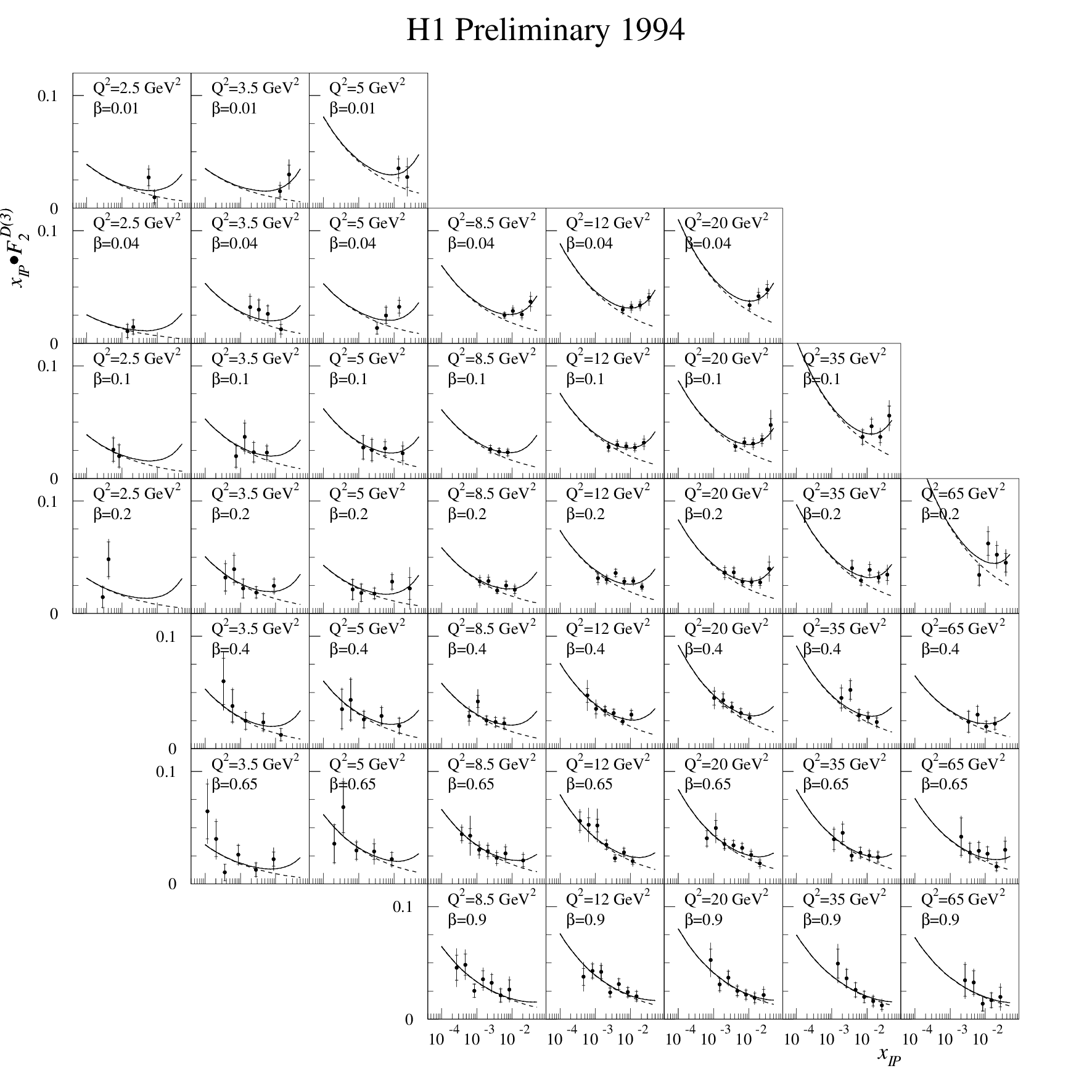}

$\;$ \vspace{18.cm}  $\;$
\bfm
\begin{center}
\leavevmode
\hbox{%
\epsfxsize = 4in
}
\end{center}
\lm{{\protect\footnotesize  ${\rm \xpom \cdot F_2^{D(2)}(\beta,Q^2)}$ together with a fit in which a pomeron component with a trajectory $\alphapom$ contributes together with a meson component with trajectory ${\rm \alpha_M}$. On each plot, the bottom line shows the contribution from the pomeron component while  the top line shows the sum of the two components. }}
\label{fig:x22}
\end{figure}

 \newpage

\par

{\bf ZEUS\_LPS determination of $\alphapom$} \nl

\noi The diffractive structure function ${\rm F_2^{D(3)}(\beta,Q^2,\xpom)}$
was  determined using LPS tagged events~\cite{lpsw}  in the range
4${\rm <Q^2<}$30 \gevtwom,  0.006$<\beta<$ 0.5, ${\rm <Q^2>=}$ 12 GeV$^2$,  
${\rm 4\cdot 10^{-4}<\xpom< 3 \cdot 10^{-2}}$ and 0.07 $<|t|<$ 0.36 GeV$^2 $, 
extending the range to lower $\beta$ and higher $\xpom$ compared
to previous ZEUS measurements (Fig.~\ref{fig:f2_roma}) .
The results are consistent with factorizable $\xpom$ dependence
in all $\beta$ bins.
Fitting the highest 3 $\beta$ bins with the same exponent gives   a=   1.28 $\pm$ 0.07 (stat.) $\pm$ 0.15 (syst.)  and therefore: 
$${\rm \alphapombar^{LPS}\; = \;  1.14\; \pm \;0.04 \;(stat.)\; \pm\; 0.08\; (syst).} $$

\bfm
\begin{center}
\leavevmode
\hbox{%
\epsfxsize = 3in
\epsffile{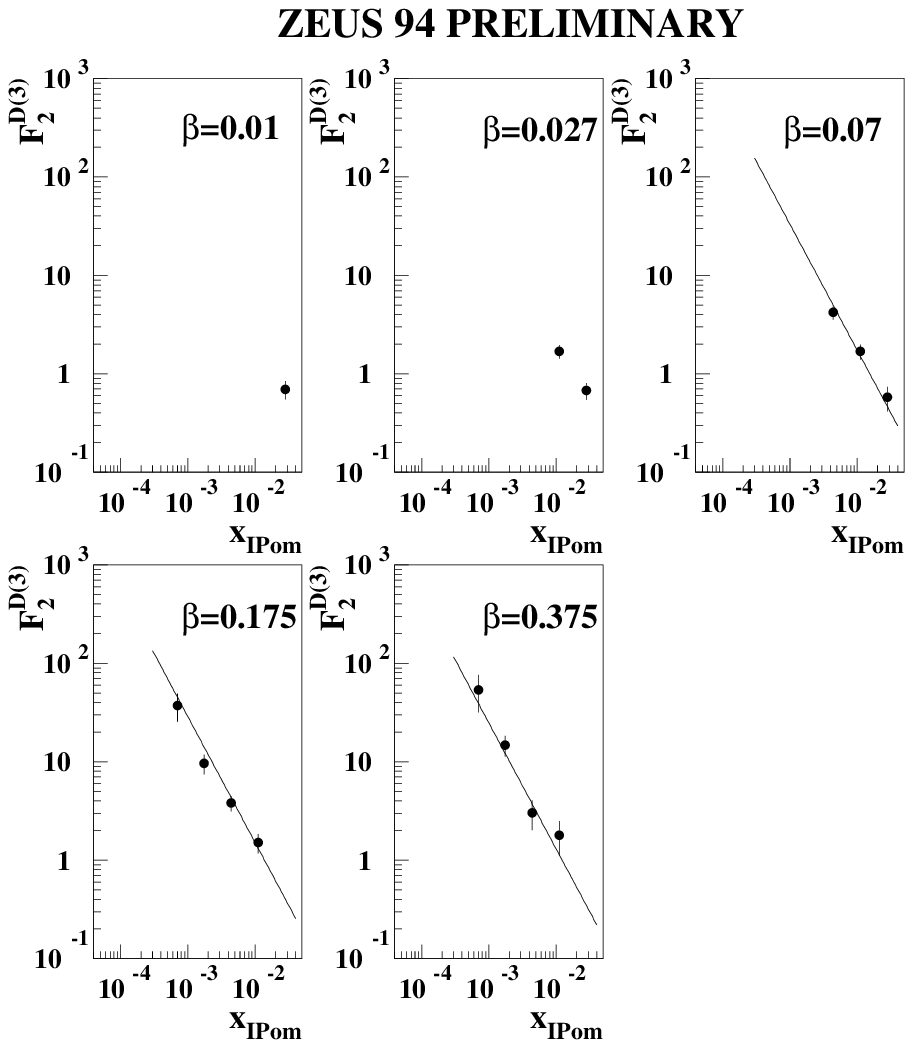}
}
\end{center}
\lm{{\protect\footnotesize The structure function  ${\rm  F_2^{D(2)}(\beta,Q^2)}$ plotted vs \xpom in bins of $\beta$ at ${\rm < Q^2 > = 12}$ \gevtwom. The errors are statistical only. The solid line corresponds to the fit described in the text.}}
\label{fig:f2_roma}
\end{figure}


\par
{\bf ZEUS\_Mx determination of $\alphapom$} \nl

\noi The determination of $\alphapom$ can also be achieved by fitting the energy dependence of the cross section in bins of \mx~\cite{ref:zeusmx} . 
In a Regge - type description ~\cite{Mueller,Fiefox} ,
the \Wgpm dependence of the diffractive cross section is of the form
\begin{eqnarray}{\rm
\frac {d\sigma_{diff}^{\gamma^*p \to XN}(\mx,\Wgpm,Q^2,t)}{ dt d\mx }  \propto (\Wgpm^2)^{2\alphapom(0) -2} \; \cdot \; 
e^{-|t|(b_0+2\alphapom^\prime\ln (W_{\gamma p}^2/(M_X^2+Q^2)))}\; \; \; ,} 
\label{eq:gxp4}
\end{eqnarray}
where ${\rm \alphapom(t) = \alphapom(0) + \alphapom^\prime t}$ is the pomeron trajectory and ${\rm b_0}$ and $\alphapom^\prime $ are  parameters.
The cross sections in each (${\rm \mx, Q^2}$) interval is fitted to the form
\begin{eqnarray}{\rm
\frac {d\sigma_{diff}^{\gamma^*p \to XN}(\mx,\Wgpm,Q^2)}{ d\mx }  \propto (\Wgpm^2)^{(2\overline{\alphapom} -2)} \; \; \; , 
}
\label{eq:gxpint}
\end{eqnarray}
allowing a determination of ${\rm \alphapombar}$. Howewer, the result obtained with this method is currently under further investigation and is shown only for completeness \footnote{ In the preliminary analysis of the ZEUS 94 data on the diffractive DIS
cross-sections a technical mistake has been found in the generation of
the Monte Carlo data used for the acceptance correction and resolution
unfolding.  This mistake led to the mishandling of QED radiative
corrections.  Its effect is to change the cross sections by typically
one systematic error.  The ZEUS collaboration thus has retracted their
1994 preliminary results until further analysis is completed and the effect on the above value of ${\rm  \alphapombar^{M_X}}$ is currently under study.}
:
$${\rm  \alphapombar^{M_X}\; \; = \; \; 1.23 \pm 0.02 (stat) \pm 0.04 (syst). }$$

\subsubsection{Comparison of the results}

\noi  
In order to compare results obtained at ${\rm Q^2 \sim 0 }$ GeV with results obtained al large \qtwom, the influence on the result of the unknown value of ${\rm \alphapom^\prime}$  in photon diffraction at large \qtwom   needs to be evaluated. Note that the  value of ${\rm \alphapom^\prime}$ in  photon diffraction at large \qtwom does not have to be the same one  measured in vector meson production in the presence of a hard scale.  Let's consider, as an example, the ln\mx analysis. Integrating over t eq.\ref{eq:gxp4}, the expression for the cross section is:
$${\rm
\frac {d\sigma_{diff}^{\gamma^*p \to XN}(\mx,\Wgpm,Q^2)}{ d\mx }  \propto (\Wgpm^2)^{2\alphapom(0) -2} \; \cdot \frac{1}{b_0+2\alphapom^\prime\ln (\Wgpm^2/(\mxtwo+Q^2))}}$$
\be{\rm 
e^{-|t|(b_0+2\alphapom^\prime\ln (W_{\gamma p}^2/(M_X^2+Q^2)))}\left |^{ |t_{min}|}_{|t_{max}|}\right .
\label{eq:inmx}
}.\ee

\noi If ${\rm \alphapom^\prime \sim 0.}$ \gevmtwom, then eq.\ref{eq:inmx} simplifies to
$$ {\rm
\frac {d\sigma_{diff}^{\gamma^*p \to XN}(\mx,\Wgpm,Q^2)}{ d\mx }  \propto (\Wgpm^2)^{2\alphapom(0) -2} },$$ 
and, comparing this expression with eq.\ref{eq:gxpint}, we obtain ${\rm \alphapombar = \alphapom(0)}$.

\noi On the other hand, if ${\rm \alphapom^\prime > 0.}$ \gevmtwom, then two effects change the slope of the \Wgpm dependence:
\bi
\item [1)] The denominator of eq.\ref{eq:inmx} is a slowly rising function of \Wgpm and therefore causes  $\alphapombar$  to be smaller than $\alphapom$. This effect has been extimated, for ${\rm \alphapom^{\prime}}$ = 0.25 \gevmtwom, to be 0.025-0.03 .

\item [2)] If the t range is limited, the last term  of   eq.\ref{eq:inmx} is a decreasing function of \Wgpm, causing   $ \alphapombar $  to be smaller than $\alphapom $.
For the ZEUS\_LPS analysis, where  ${\rm |t_{min}| = 0.07 \; GeV^2}$,  a value of  ${\rm \alphapom^{\prime}}$ = 0.25 \gevmtwom $\;$ reduces the measured $\alphapombar$ value by  $ \sim$ 0.02 .

\ei

\noi In Fig.~\ref{fig:alphapom}, the compilation of $\alphapom$ values obtained at HERA is shown assuming, for the measurement at large \qtwom, (a) $\alphapom^\prime$ = 0 \gevmtwom $ \;$ ; or (b)   $\alphapom^\prime$ = 0.25 \gevmtwom $\;$ (b).
 The solid line is the statistical error while the dotted line is the systematic error. The values indicated as H1 93 (${\rm \eta_{max}}$)~\cite{ref:h1eta} and ZEUS 93 (${\rm \eta_{max}}$)~\cite{ref:zeuseta} are the first measurement obtained by each experiment and they were obtained using the selection cut ${\rm \eta_{max} < 1.8}$ for H1 and  ${\rm \eta_{max} < 1.5 }$ for ZEUS.

\bfm
\begin{center}
\leavevmode
\hbox{%
\epsfxsize = 5in
\epsffile{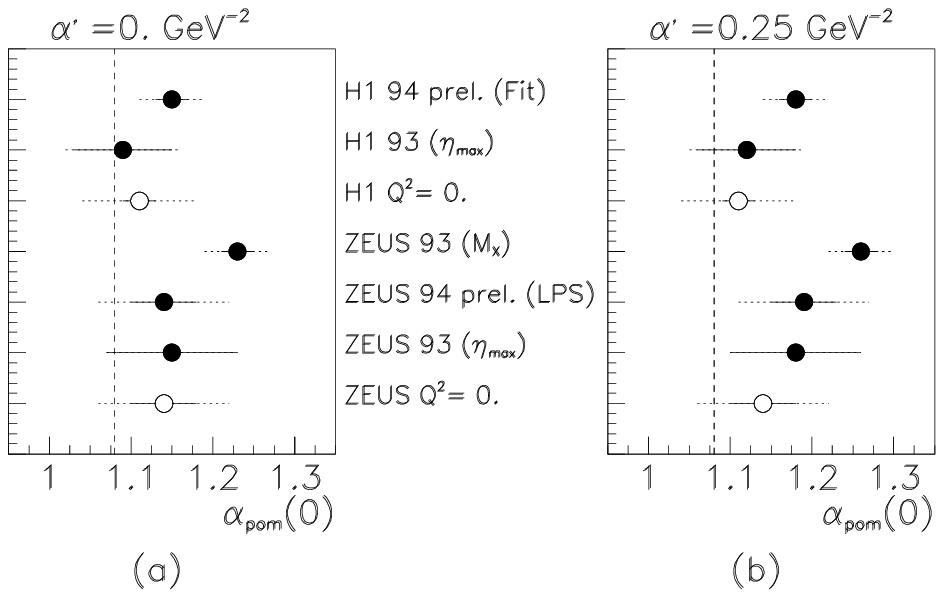}
}
\end{center}
\lm{{\protect\footnotesize
Compilation of ${\rm \alphapom(0)}$  values obtained at HERA,  assuming, for the measurement at large \qtwom,  (a) $\alphapom^\prime$ = 0. \gevmtwom $ \;$ ; or (b)   $\alphapom^\prime$ = 0.25 \gevmtwom $\;$ (b). Empty dots are values obtained at \qtwom = 0. \gevtwom, full dots at high \qtwom. The dashed vertical line  is the value ${\rm \alphapom(0)}$ = 1.08. The solid line is the statistical error while the dotted line is the systematic error.
}}
\label{fig:alphapom}
\end{figure}


\noi It is possible that the  difference in value between ${\rm {\alphapom}^{Q^2=0}(0)}$ and ${\rm {\alphapom}^{Q^2>0}(0)}$ is a signal for the presence of a small `hard' pomeron component in the diffractive sample at high \qtwom. How to measure it, its magnitude  and how to enhance it choosing particular final states (for example see~\cite{ref:land,ref:bartqq}) is currently under intense theoretical investigation. Note also that the above comparison is done among measurements performed on different t and \xpom ranges.

\subsection{Measurement of the slope parameter b in diffractive DIS}

Using the ZEUS leading proton spectrometer, the t distribution of diffractive DIS was measured directly for the first time at HERA~\cite{lpsw} .
 The measurement of t has been performed in the kinematic range:
 ${\rm x_L> 0.97,\; 4 <Q^2< 30\; GeV^2,\; }$ ${\rm <Q^2>=\; 12 GeV^2}$, 70 $<\Wgpm<$ 210 GeV,
0.07 ${\rm <|t|<}$ 0.36 GeV$^2$. Assuming  an exponential behaviour ${\rm\frac{d\sigma}{dt}\propto e^{-b|t|}}$, b is measured to be:
$${\rm b\;=\; 5.9\; \pm\; 1.3\; (stat.)\;^{+1.1}_{-0.7}\;(syst) GeV^{-2}}.$$
The measured t distribution is shown in Fig.~\ref{fig:ts}. The value of b is similar to the values obtained in single diffraction in pp interactions.

\bfm
\begin{center}
\leavevmode
\hbox{%
\epsfxsize = 3in
\epsffile{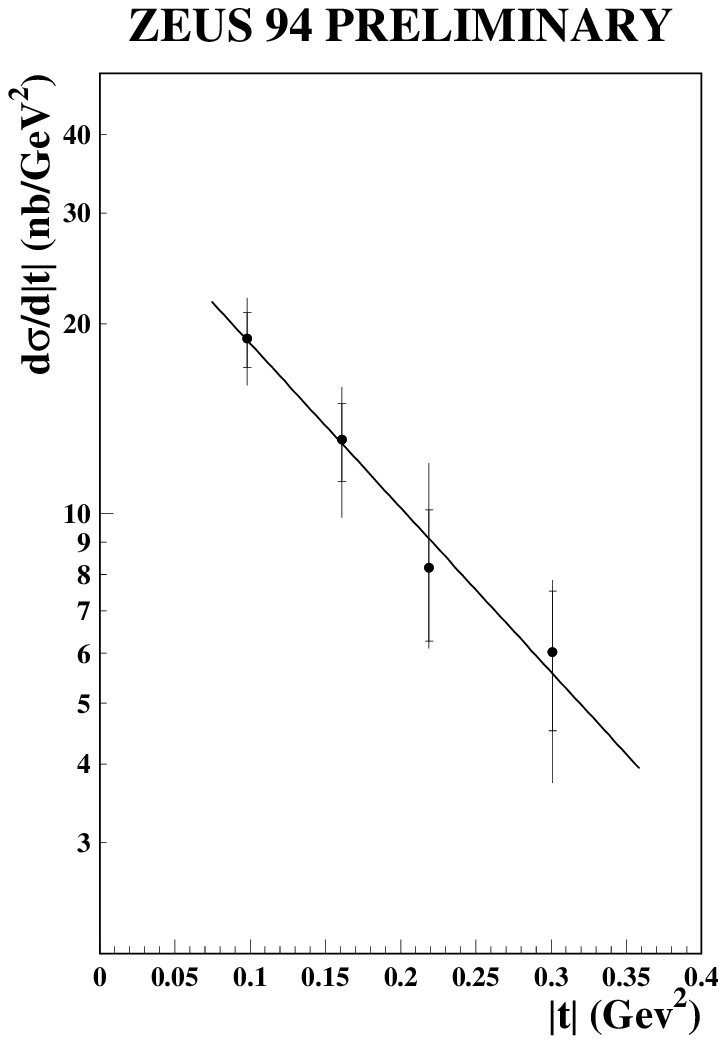}
}
\end{center}
\vspace{-1.5cm}
\lm{\protect\footnotesize Differential cross section ${\rm \frac{d\sigma}{dt}}$ for diffractive
DIS events with a leading proton  detected in the LPS.
The error bars represent the statistical and systematic errors added in
quadrature and the line shows the result of the exponential fit.}
\label{fig:ts}
\end{figure}
\vspace{-1.cm}
\subsection{Partonic structure of diffractive exchange}

\subsubsection{QCD fit to the diffractive structure function}
The H1 collaboration~\cite{ref:hf2d} performed a QCD analysis of the diffractive structure function \ftwodthree.  The analysis is performed integrating \ftwodthree $\;$ over the measured \xpom range and the result  interpreted as the deep inelastic structure of the exchanged object averaged over t and \xpom:
 
$${\rm  \tilde{F}_2^{\pom}(Q^2,\beta) = \int_{\xpom_{min}}^{\xpom_{max}}
(\frac{1}{\xpom})^n \cdot  F_2^{\pom}(Q^2,\beta)\cdot d\xpom}.
$$
The QCD analysis is performed fitting the data using a  flavour singlet quark and gluon distribution   (${\rm u+\bar{u}+d+\bar{d}+s+\bar{s}\; +\; gluon }$) at a starting scale ${\rm Q_0^2=2.5\,{\rm GeV}^2}$ and then evolving the system according to the DGLAP~\cite{ref:dglap} evolution equation. The results are shown in Fig.~\ref{fig:qcdboth_new}. The most striking feature in the data is that a rise with ln\qtwom persists to values of $\beta$ far in excess of the point (\xbj $\simeq$ 0.15) at which the structure of the proton is dominated by  quarks rather than by gluons suggesting a strong gluonic component in the structure of the diffractive exchange. The QCD fit supports this interpretation: the analysis has been also done  considering only quarks at the starting scale ${\rm Q_0^2}$ and  a much worst $\chi^2$  value ha been obtained. The parton distributions obtained from the fit are shown in  Fig.~\ref{fig:qcd_comb}. At \qtwom = 5 \gevtwom $\;$ a `leading' gluon behaviour is observed, in which the exchange is dominated by gluons carrying a very large fraction of the longitudinal momentum.

\bfm
\begin{center}
\leavevmode
\hbox{%
\epsfxsize = 4in
\epsffile{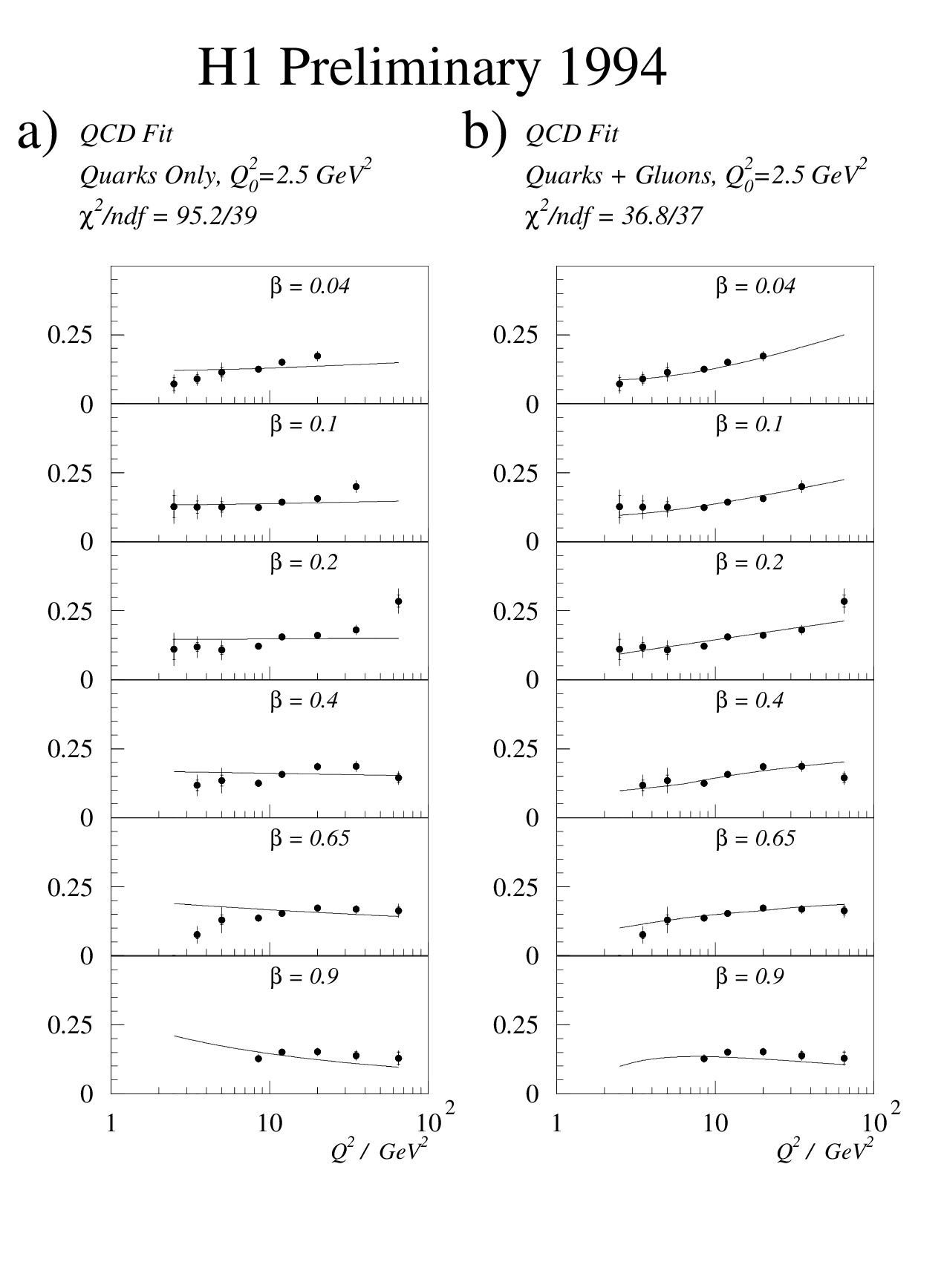}}
\end{center}
\vspace{-1.cm}
\lm{{\protect\footnotesize  DGLAP QCD comparison of the ${\rm (\beta, Q^2)}$ dependence of ${\rm \tilde{F_2^D}}$: a) assuming only quarks at the starting scale of ${\rm Q^2_0 = 2.5}$ \gevtwom, b) assuming both quarks and gluons  at the starting scale of ${\rm Q^2_o = 2.5}$ \gevtwom. }}
\label{fig:qcdboth_new}
\end{figure}

\bfm
\begin{center}
\leavevmode
\hbox{%
\epsfxsize = 4in
\epsffile{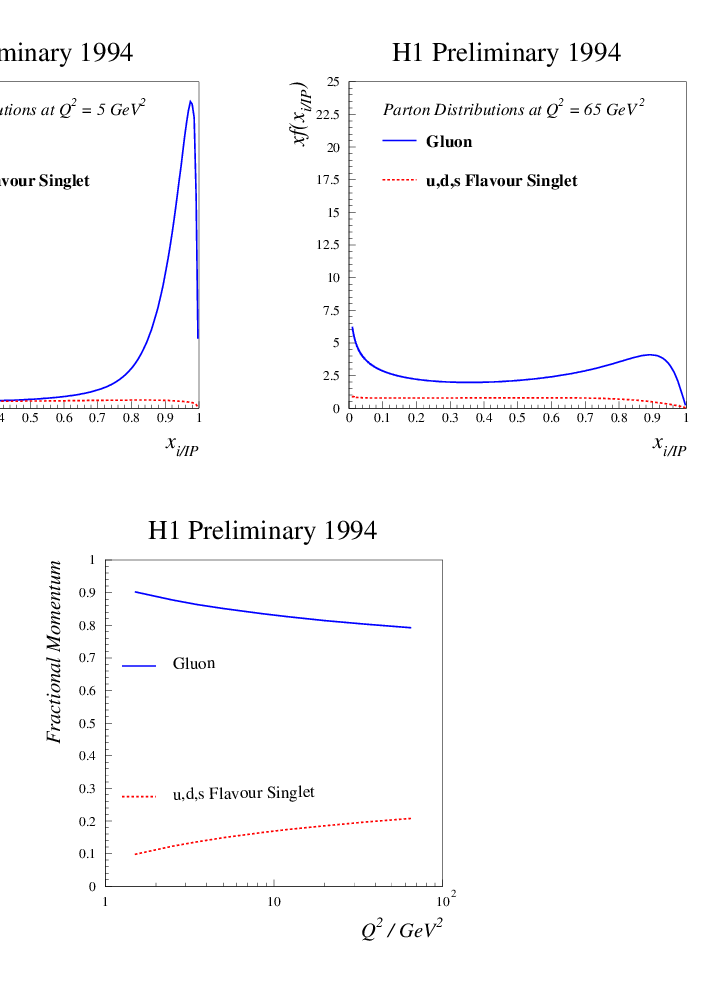}}
\end{center}
\lm{{\protect\footnotesize Quark and gluon  fractional momentum distributions for diffractive exchange averaged over \xpom and t extracted using the DGLAP QCD fit at a) \qtwom = 5 \gevtwom and b)  \qtwom = 65 \gevtwom; c) fraction of the total momentum carried by quarks and by gluons as a function of \qtwom.  }}
\label{fig:qcd_comb}
\end{figure}

\newpage
\subsubsection{Jet structure}
The question of the constituent content of the pomeron can also be addressed via measurements of diffractively produced jets, both in photoproduction~\cite{ref:zjphp,ref:zjw} and DIS~\cite{ref:hjw} . The ZEUS collaboration studied the diffractive dijet cross section:
$${\rm \gamma p \longrightarrow jet + jet + X +p }$$
and compared its magnitude and shape with different model predictions based on a factorizable model of pomeron. To ensure diffractive production, a gap in the most forward part of the detector of at least 2.9 unit was required.

\noi  The following Pomeron fractional momentum densities were used in the MC: \\
Super-hard gluon:  ${\rm \beta f_{g / I\!\!P}(\beta) = \frac{0.1}{(1-\beta)^{0.9}}}$\\
Hard gluon: ${\rm \beta f_{g / I\!\!P}(\beta) = 6 \beta (1 - \beta) \;\;,<\beta>=1/2}$\\
Hard-quark (2 flavours): ${\rm \beta f_{q / I\!\!P}(\beta) = \frac{6}{4} \beta (1 - \beta)}$.

\noi The measured distribution, Fig.~\ref{fig:dijet}, is compatible with a Pomeron containing a hard-gluon density.

\bfm
\begin{center}
\leavevmode
\hbox{%
\epsfxsize = 3in
\epsffile{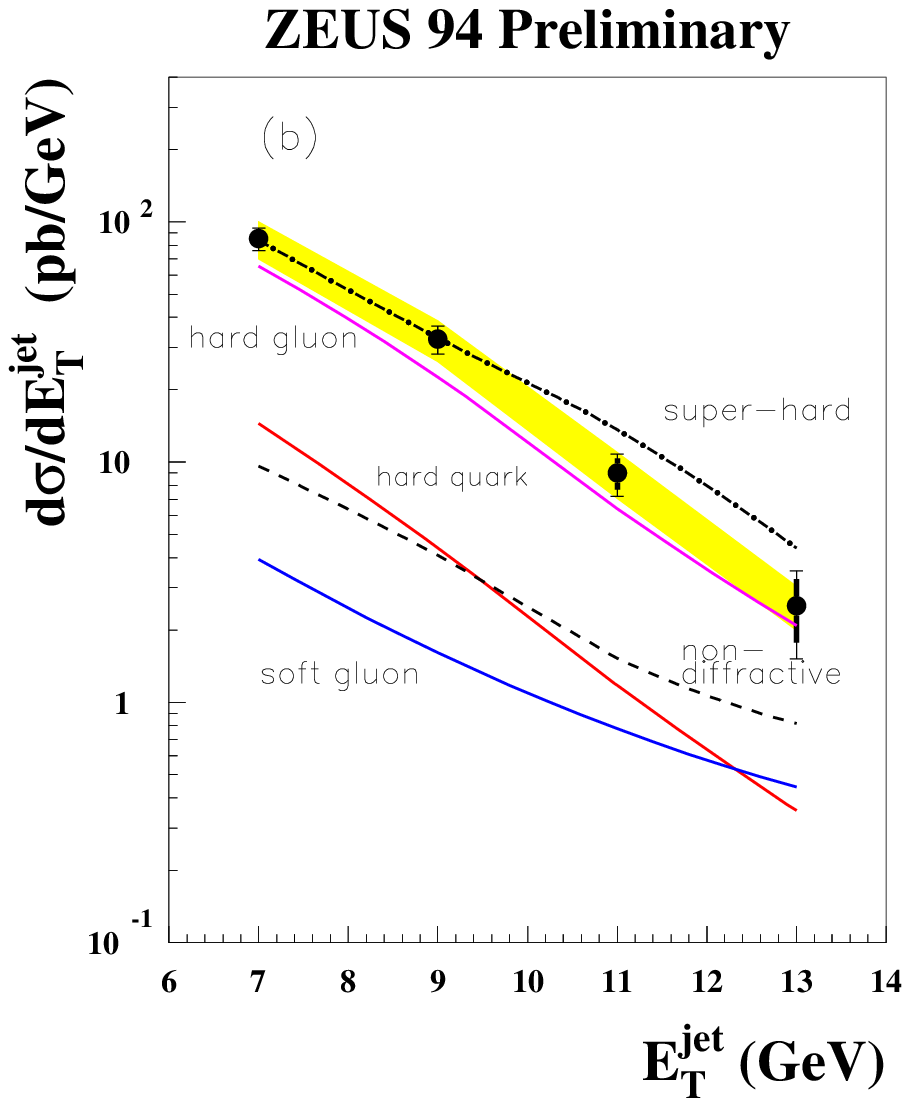}
}
\end{center}
\lm{{\protect\footnotesize The diffractive dijet cross section as a function of ${\rm \eta^{jet}}$ compared to MC predictions for different pomeron momentum densities.}}
\label{fig:dijet}
\end{figure}
\noi  The thick error bars represent the statistical errors
of the data and the thin error bars show the statistical  error added in quadrature with the  systematic non associated with the jet energy scale.
Comparison with POMPYT~\cite{ref:pompyt} MC calculations for a gluonic (quarkonic)
Pomeron are shown, including both direct and resolved contributions
and different choices of the parton density. From top to the bottom
the curves correspond to super-hard gluon (dashed-dotted), hard gluon,
hard quark and soft gluon.
The non-diffractive
contribution modelled by PYTHIA~\cite{pythia} is shown as a dashed line.

\subsection{${\rm \sigma_{dif}/\sigma_{tot}}$ as a function of \qtwom}
 Fig.~\ref{fig:rat} shows the ratio between the diffractive and total cross section as a function of \qtwom. At \qtwom $\sim$ 0 \gevtwom, the diffractive part of the cross section is  ${\rm 36 \pm 8\%}$ (${\rm 42 \pm 8\%}$)  according to the ZEUS (H1) collaboration  of which  ${\rm 23 \pm 6 }$\%  (${\rm 32\pm4}$ \% ) is photon diffraction and 13$\pm$5 \% (10$\pm$3 \%) is vector meson production. 
This last component  has been measured to fall at least like ${\rm d\sigma/dQ^2 \sim 1/(m^2_{VM}+Q^2)^2}$ and therefore becomes negligible as \qtwom increases. Photon diffraction seems to decrease going from  \qtwom = 0 to 10 \gevtwom $\;$ while it stays flat as a function of \qtwom at large \qtwom indicating that diffraction is a leading twist mechanism. It will be very interesting to have more accurate data to see if this decrease, assuming that it is actually confirmed,  happens at the same \qtwom values  where pQCD starts to be applicable, indicating a change in the nature of the ${\rm \gamma p}$ diffractive interaction.

\bfm
\begin{center}
\leavevmode
\hbox{%
\epsfxsize = 3in
\epsffile{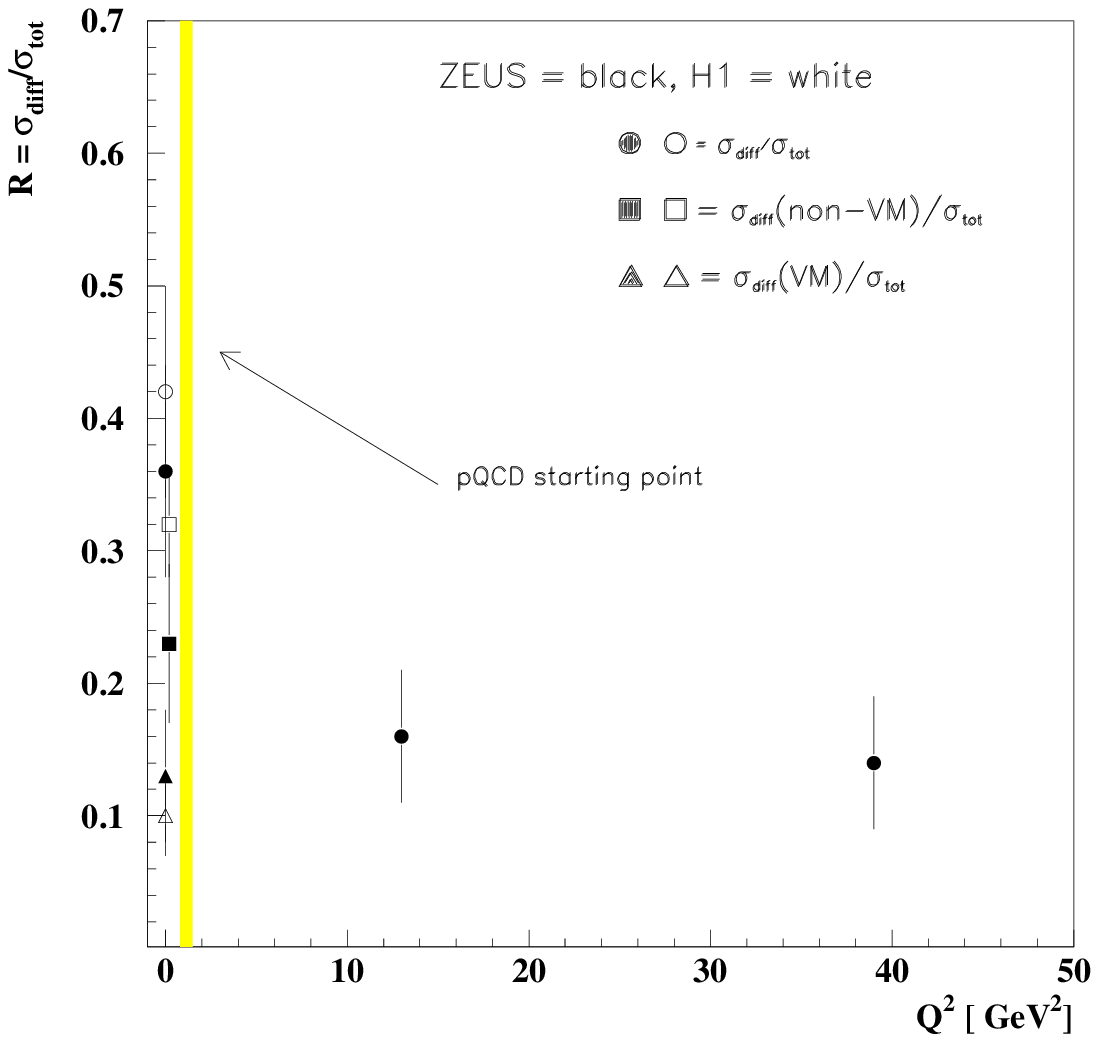}
}
\end{center}
\lm{{\protect\footnotesize  Ratio between the diffractive and total cross section as a function of \qtwom.
}}
\label{fig:rat}
\end{figure}

\section{Central rapidity gaps } 

\noi In high energy hadronic collisions, the dominant mechanism for jet production
is  a hard scatter between partons in the incoming hadrons via a
quark or gluon propagator. 
Such jets are said to be
`colour connected' and this leads to the production of particles throughout
the rapidity region between the jets.  However, if the hard scattering were
mediated by the exchange of a colour singlet propagator in the t-channel,
each jet would be colour connected only to the beam remnant closest in
rapidity and the rapidity region between the jets would contain few
final-state particles~\cite{dokshitzer} , Fig.~\protect\ref{fig:diagramcs}(a,b)

D0~\cite{D0} and CDF~\cite{CDF} have reported the results of searches at ${\rm \sqrt{s} = 1.8}$~TeV for dijet events in
${\rm p\bar{p}}$ collisions containing a
rapidity gap between the two highest transverse energy (${\rm E_T^{jet}}$) jets. 
Both collaborations see an excess of gap events over the expectations from
colour exchange processes. D0 reports an excess of ${\rm 0.0107 \pm 0.0010(stat.)
^{+0.0025} _{-0.0013}(syst.)}$, whereas CDF measures the fraction to be ${\rm 0.0086
\pm 0.0012}$. 

At HERA an equivalent mechanism is possible, with the hadronic fluctuation
of the photon acting as one of the hadrons. 
In order to quantify the rapidity gap events,
a gap-fraction, ${\rm f(\Delta\eta)}$, is defined as the ratio of the number of
dijet events  which have a rapidity gap of width \deta\ between the jets to the
total number of dijet events. As explained above, for colour non-singlet exchange,
the gap-fraction is expected to fall exponentially with increasing \deta while for colour singlet exchange, the gap-fraction is
not expected to depend strongly upon ${\rm \Delta\eta}$~\cite{ref:bj,ref:deld}.
The situation is illustrated in  Fig.~\protect\ref{fig:diagramcs}(d).


\bfm
\begin{center}
\leavevmode
\hbox{%
\epsfxsize = 4.in
\epsffile{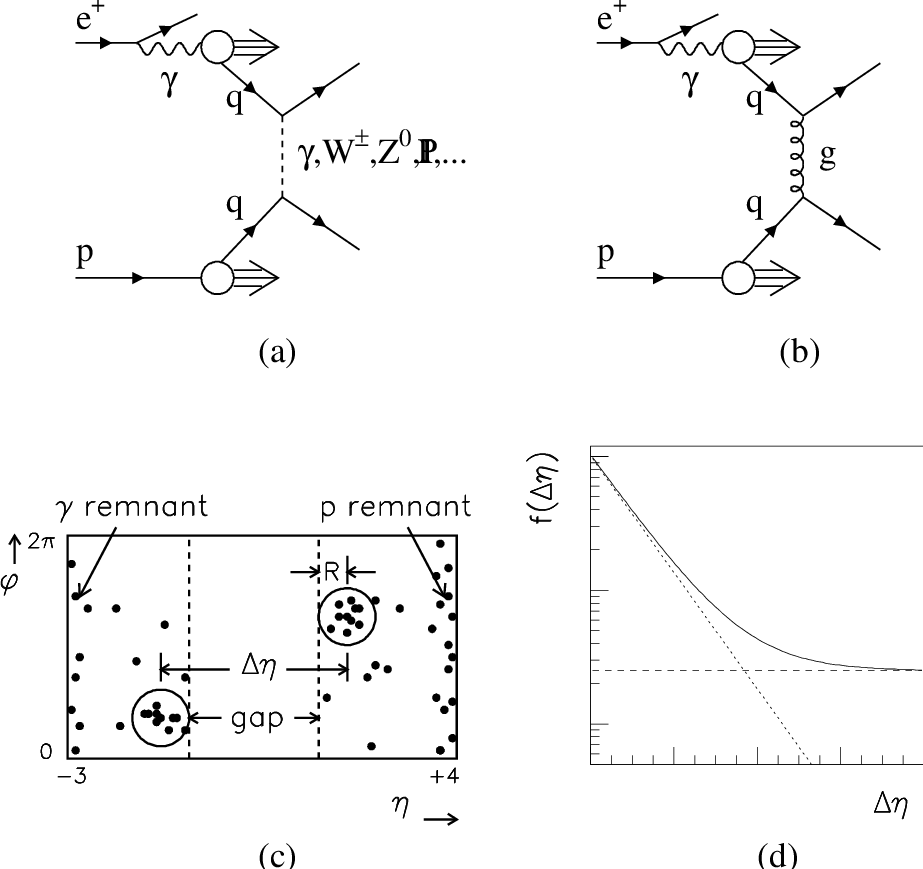}
}
\caption{\protect\footnotesize
Resolved photoproduction via (a) colour singlet exchange and (b) 
colour non-singlet exchange.
The rapidity gap event morphology is shown in (c) where black dots
represent final state hadrons and the boundary illustrates the limit
of the ZEUS acceptance.  Two jets of radius R are shown, which are back
to back in azimuth and separated by a pseudorapidity interval
${\rm \Delta\eta}$.
An expectation for the behaviour of the gap fraction is shown
in (d)(solid line).
The non-singlet contribution is shown as the dotted line and 
the colour singlet contribution as the dashed line.}
\label{fig:diagramcs}
\end{center}
\end{figure}

\bfm
\begin{center}
\leavevmode
\hbox{%
\epsfxsize = 4in
\epsffile{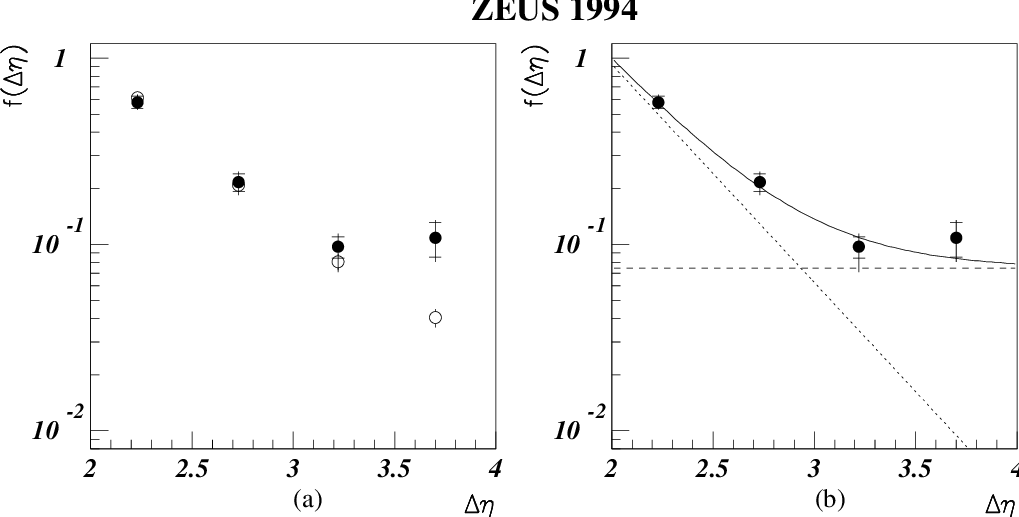}
}
\end{center}
\lm{{\protect\footnotesize Results from events with central rapidity gaps,
(a) Points before (open circles) and after (full circles) detector corrections.
(b) Fit as explained in the text. }}
\label{fig:cgap3}
\end{figure}
The results~\cite{ref:zrg} are shown in  Fig.~\protect\ref{fig:cgap3}. Both the comparison with the default PYTHIA non-singlet prediction and the
fit to an exponential form give an excess of about 0.07 in the gap-fraction
over the expectation from colour non-singlet exchange. This excess can be interpreted as evidence of hard diffraction: a
simple two-gluon model for pomeron exchange gives ${\rm \hat{f}(\Delta\eta) \sim
0.1}$~\cite{ref:bj} thus showing that pomeron exchange could account for the data.

The magnitude of the squared four-momentum transfer across the rapidity gap
as calculated from the jets is large (${\rm |t| \ge (E_T^{jet})^2}$).  Thus the
colour singlet exchange is unambiguously `hard'. 

\section{Conclusions}

Diffraction at HERA has provided many measurements in both the soft and perturbative domains. 

\noi - The rise of the  total ${\rm {\gamma p}}$ cross section  has been measured to be weak, consistent with the exchange of the same `soft' pomeron responsible for the rise with energy of hadronic reactions.

\noi
- Diffractive photoproduction is also  governed by soft pomeron exchange: a value of ${\rm \alphapom(0) \simeq 1.11-1.14}$ has been measured from the mass spectrum of the dissociated photon in the triple pomeron regime.

\noi - Elastic vector meson production  at HERA  shows  a clear distinction between two classes of processes. A first group of results, photoproduction of light vector mesons (${\rm \gamma p \rightarrow Vp,\; V= \rho^0,\;\phi \;\omega}$),  exhibits the characteristic features of diffraction, as described by Regge theory: a weak energy dependence of the cross section and a value of the t slope parameter b as observed in hadronic diffractive reactions. They are therefore explained in term of the same pomeron that controls the total cross section. A second group, which includes photoproduction of $J/\psi$ and light vector meson production at high \qtwom shows a different pattern: a  strong energy dependence of the cross section, a values of b consistent with a point like $\gamma$ V  vertex and the restoration of the SU(4) flavour symmetry indicate a type of dynamic consistent with pQCD predictions.

\noi - Photon diffraction at large \qtwom shows a value of $\alphapom$ only slightly higher than the values obtained in photoproduction, indicating that the same mechanism  used to explain photoproduction processes can be used to explain a large fraction of diffractive dissociation at high \qtwom.  The partonic structure of the  pomeron has been measured and found to be dominated by hard gluons. Factorization has been found to be valid within the current sensitivity and measurements range.

\section{Acknowledgements}

It is a pleasure to thank in H. Abramovich, M. Arneodo, J. Dainton, E. Gallo, G. Iacobucci, A. Levi, R. Nania, J. Phillips,  F. Sciulli, A. Solano and to Lilian DePorcel for her infinite patience and understanding.



\begin{thebibliography}{99}


\bibitem{h1} H1 Collaboration, DESY 93-103. 
\bibitem{zeus} ZEUS Collaboration, The ZEUS detector, Status Report (1993). 
\bibitem{ref:peppe} G. Iacobucci,  Talk given at  the 1996 Zeuthen Workshop on Elementary  Particle Theory, Rheinsberg, Germany (1996), \\
A. Staiano, Talk given at `Les Rencontres de la Physique de la Valle d'Aoste' La Thuile (1996),    \\
G. Barbagli,  Talk given at  the 1996 IHEP Conference, Warsaw.
\bibitem{ref:col}  P.D.B. ~Collins, "An Introduction to Regge Theory and
 High Energy Physics",
Cambridge University Press, Cambridge (1977).
\bibitem{ref:pearl} M. Perl, High Energy Hadron Physics,\\
 Wiley \& Son, New York, 1974.
\bibitem{ref:donlan}A.Donnachie, P.V.Landshoff, \pl {\bf B296} (1992) 227. 
\bibitem{ref:h1tot} \hcola, \zp {\bf C69} (1995) 27.
\bibitem{ref:ztot} \zcol, \zp {\bf C63} (1994) 391.
\bibitem{ref:cdftcs}CDF Coll., F.Abe et al., \prev {\bf D50} (1994) 5550. 
\bibitem{ref:hal} H. Abramowicz et al., \pl {\bf B269} (1991) 465.
\bibitem{ref:zf2}\zcol,  DESY 96-076 (June 1996). 
\bibitem{ref:hf2} \hcol,  \np {\bf B470} (1996).
\bibitem{ref:ryb} M. Ryskin et al.,`Heavy Photon Dissociation in DIS', Proceeding of the workshop `Physics at HERA', Hamburg (1991).
\bibitem{ref:kop} B. Kopeliovich et al, hep-ph/9601291.
\bibitem{ref:bj} J. D. Bjorken, \prev {\bf D47} (1992) 101.
\bibitem{ref:ing} M. Przybycien et al., hep-ph/9606294.
\bibitem{don84}A.~Donnachie, P.V.~Landshoff, \np  {\bf B244} (1984) 322.
\bibitem{don87}A.~Donnachie, P.V.~Landshoff, \pl {\bf B191} (1987) 309.
\bibitem{ingsch} G. Ingelman and P. Schlein, \pl {\bf B152} (1985) 256.

\bibitem{nik91}N.N.~Nikolaev, B.G~Zakharov, \pl {\bf B260} (1991) 414.
\bibitem{abr} H. Abramowicz et al. , SLAC Summer Inst. (1994) 539. 
\bibitem{ref:nz2} N. Nikolaev et al., \zp {\bf C53} (1992) 331.
\bibitem{gots} E. Gotsman et al., hep-ph/9606280.
\bibitem{levin}E. Levin et al., \prev {\bf D50} (1994) 4306.
\bibitem{genovese} M. Genovese et al., \spj  {\bf 81} (1995) 625.
\bibitem{ref:nz3}  M. Genovese et al., \pl {\bf B380} (1996) 213.
\bibitem{ref:alj} J. Bjorken,AIP Conference Proceedings No. 6, Particles and Fields Subseries N0. 2, Ed. M. Bander, G. Shaw and D. Wong (AIP, New York,1972) \\
J. D. Bjorken and J. Kogut, \prev {\bf D8} (1973) 1341, \\
J. D. Bjorken, preprint SLAC-PUB-7096 (1996), hep-ph/9601363.
\bibitem{ref:buch} W. Buchmuller et al., \pl {\bf B355} (1995) 573.
\bibitem{ref:fey} R. Feynman, Photon-Hadron Interactions, Addison-Wesley, 1989.
\bibitem{ref:ry} M.G. Ryskin, \zp {\bf C57} (1993) 89. 
\bibitem{ref:br} S.J. Brodsky et al., \prev  {\bf D50} (1994) 3134. 
\bibitem{ref:ne} J. Nemchik et al., \pl  {\bf B341} (1994) 228. 
\bibitem{ref:ry2} A.D. Martin et al. hep-ph/9609448
\bibitem{ref:zrho2}\zcol, \pl {\bf B356} (1995) 601.
\bibitem{ref:hrho1} \hcol,  \np {\bf B463} (1996) 3.
\bibitem{ref:zrho3}\zcol, DESY 96-183  , accepted by Zeitschrift f. Physik - MS 418
\bibitem{ref:zrho1}\zcol, \zp {\bf C69} (1995) 39.
\bibitem{ref:hrhoj1} \hcol, \np {\bf B468} (1996) 3.
\bibitem{ref:zphi1}\zcol, \pl {\bf B377} (1996) 259.
\bibitem{ref:zphi2}\zcol,\pl {\bf B380} (1996) 220.
\bibitem{ref:hjp1} \hcol, \pl {\bf  B338} (1994) 507.  
\bibitem{ref:zjp1}\zcol, \pl {\bf B350} (1995) 120.
\bibitem{ref:hjp2}\hcol, \np {\bf B472} (1996) 3. 
\bibitem{ref:zw1}\zcol, ICHEP 1996, pa 02-051.
\bibitem{ref:zw2}\zcol, ICHEP 1996, pa 02-053.
\bibitem{ref:zw3}\zcol, ICHEP 1996, pa 02-028.
\bibitem{ref:hw2}\hcol, ICHEP 1996, pa 02-064.
\bibitem{ref:hw3}\hcol, ICHEP 1996, pa 02-085.
\bibitem{ref:zw4}\zcol, ICHEP 1996, pa 02-049.
\bibitem{ref:jp2s1}\hcol, ICHEP 1996, pa 02-086.
\bibitem{ref:rhop1}\hcol, ICHEP 1996, pa 01-088.
\bibitem{ref:cdfdd}CDF Collaboration, F. Abe, et al., \prev  {\bf D50}, (1994) 5535
(1995) 855.
\bibitem{ref:curv} A. Schiz et al., \prev {\bf D24} (1981) 26.
\bibitem{ref:lps}A.Staiano, Silicon Detectors for the Leading Proton Spectrometer of ZEUS, 
                 proceedings of the  Third~International~Workshop~on~Vertex~Detectors , IUHEE-95-1;\\
                 K. O'Shaughnessy et al.,  Nucl.~Instrum.~Methods  {\bf A342}, (1994) 260-263.
\bibitem{ref:frank} L. Frankfurt et al., \prev {\bf D54} (1996) 3194.
\bibitem{ref:nzsh}N. N.  Nikolaev et al.,  \pl {\bf B366}  (1996) 337.
\bibitem{ref:hal2} H. Abramowicz et al, Proceeding of `Future Physics at HERA', Hamburg, 1996.
\bibitem{ref:zrap1} \zcol,  \pl {\bf B315} (1993) 481.
\bibitem{ref:zeusmx}\zcol, \zp {\bf C70} (1996) 391.
\bibitem{ref:gou} K.Goulianos, \prep {\bf 101} (1983) 169.
\bibitem{ref:hmx}  \hcol, ICHEP 1996, pa 02-067.
\bibitem{ref:zmx}\zcol, ICHEP 1996, pa 02-048.
\bibitem{ref:zrap2} \zcol, \zp {\bf C67} (1995) 227.
\bibitem{ref:hf2d} \hcol, ICHEP 1996, pa 02-061.
\bibitem{lpsw}\zcol, ICHEP 1996, pa02-026.
\bibitem{Mueller} A.H. Mueller, \prev {\bf D2}~(1970)~2963; ibid. 
{\bf D4}~(1971) 150.  
\bibitem{Fiefox} R.D. Field and G. Fox, \np {\bf B80}~(1974)~367.
\bibitem{ref:h1eta} \hcol, \pl {\bf B348} (1995) 681.
\bibitem{ref:zeuseta} \zcol, \zp {\bf C68} (1995) 569.
\bibitem{ref:land} P.V. Landshoff, Talk given at  International Workshop on Deep Inelastic Scattering and Related Phenomena (DIS 96), Rome, Italy, 15-19 Apr 1996, hep-ph/9605331
\bibitem{ref:bartqq} J. Bartels et al., \pl {\bf B379} (1996) 239, ERRATUM-ibid. {\bf B382} (1996) 449. 
\bibitem{ref:dglap} Yu. L. Dokshitzer, JETP {\bf 46} (1977) 641 \nl
V. N. Gribov and L. N. Lipatov, Sov. J. Nucl Phys. {\bf 15} (1972) 78. \nl
G. Altarelli and G. Parisi, \np {\bf B126} (1977) 298.
\bibitem{ref:zjphp}\zcol , \pl {\bf B356} (1995) 129.
\bibitem{ref:zjw}\zcol, ICHEP 1996, pa 02-039.
\bibitem{ref:hjw} \hcol, ICHEP 1996, pa 02-068.
\bibitem{ref:pompyt} P. Bruni and G. Ingelman, DESY 93-187.
\bibitem{pythia} PYTHIA 5.6: H.-U. Bengtsson and T. Sj\"ostrand, Comp. Ph
ys.   Comm. {\bf 46}~(1987)~43.
\bibitem{dokshitzer} Y. Dokshitzer, V. Khoze and S. Troyan, in 
Proceedings of the 6th International Conference on Physics in Collisions,
Chicago, Illinois, ed. M. Derrick (World Scientific, Singapore, 1987)
417.
\bibitem{D0} D0 Collaboration, S. Abachi et al., \prl {\bf 72} (1994) 2332; \\
D0 Collaboration, S. Abachi et al., FERMILAB-PUB-95-302-E (1995). 
\bibitem{CDF} CDF Collaboration, F. Abe, et al., \prl {\bf 74} (1995) 855.
\bibitem{ref:deld} V. Del Duca and W.-K. Tang, \pl {\bf B312} (1993) 225.
\bibitem{ref:zrg} \zcol, \pl {\rb B369} (1996) 55. 

\end{thebibliography}
\end{document}